%% file: papv1.tex
\documentclass[12pt]{article}
\usepackage{amsmath}
\usepackage{amsfonts}
\usepackage{amssymb}
\usepackage{mathrsfs}
\usepackage{multicol}
\usepackage{color}
\usepackage{graphicx}
\usepackage{pstricks}
\usepackage{pst-node}
\usepackage{wrapfig}
\usepackage{enumerate}
\usepackage{caption}
\usepackage{cite}
\usepackage{subfigure}
\usepackage{amsfonts,amssymb}
\usepackage{amsthm}
\usepackage{amscd}
\definecolor{gray1}{gray}{0.1}
\definecolor{gray2}{gray}{0.2}
\definecolor{gray3}{gray}{0.3}
\definecolor{gray4}{gray}{0.4}
\definecolor{gray5}{gray}{0.5}
\definecolor{gray6}{gray}{0.6}
\definecolor{gray7}{gray}{0.7}
\definecolor{gray8}{gray}{0.8}
\definecolor{gray9}{gray}{0.9}

\newpsobject{showgrid}{psgrid}{subgriddiv=1,griddots=10,gridlabels=6pt}

\definecolor{dark-green}{rgb}{0,0.7,0}
\definecolor{dark-blue}{rgb}{0,0.2,0.5}
\definecolor{med-blue}{rgb}{0,0.7,1}
\definecolor{mblue}{rgb}{0,0.2,1}
\definecolor{cnc}{rgb}{0.8,0,0}
\definecolor{light-red}{rgb}{1,0.8,0.8}
\definecolor{dark-yellow}{rgb}{1,0.8,0}
\definecolor{light-blue}{rgb}{0.8,0.9,1}
\definecolor{verylight-blue}{rgb}{0.93,0.95,1}
\definecolor{light-yellow}{rgb}{1,0.9,0.8}
\definecolor{grey}{gray}{0.88}

\newpsobject{showgrid}{psgrid}{subgriddiv=1,griddots=10,gridlabels=6pt}

\begin{document}

\thispagestyle{empty}

\setlength{\abovecaptionskip}{10pt}

\begin{center}
{\Large\bfseries\sffamily{Spontaneous scalarization of charged black holes \\
at the approach to extremality}}
\end{center}
\vskip 3mm
\begin{center}
{\bfseries{\sffamily{Yves Brihaye$^{\rm 1}$ and Betti Hartmann$^{\rm 2}$}}}\\
\vskip 3mm{$^{\rm 1}$\normalsize Physique-Math\'ematique, Universit\'e de Mons-Hainaut, 7000 Mons, Belgium}\\
{$^{\rm 2}$\normalsize{Instituto de F\'isica de S\~ao Carlos (IFSC), Universidade de S\~ao Paulo (USP), CP 369, \\
13560-970 , S\~ao Carlos, SP, Brazil}}
\end{center}

\begin{abstract} 
We study static, spherically symmetric and electrically charged black hole solutions in a quadratic Einstein-scalar-Gauss-Bonnet gravity model.
Very similar to the uncharged case, black holes
undergo {\it spontaneous scalarization} for sufficiently large scalar-tensor coupling $\gamma$ --
a phenomenon attributed to a tachyonic instability 
of the scalar field system. While in the uncharged case, this effect is only possible for positive values of $\gamma$, we show that for sufficiently large values of the electric charge $Q$ two independent domains of existence in the $\gamma$-$Q$-plane appear~: one for positive $\gamma$ and one for negative $\gamma$. We demonstrate that this new domain for negative $\gamma$ exists because of the fact that 
the near-horizon geometry of a nearly extremally charged black hole is $AdS_2\times S^2$.  
This new domain appears for electric charges larger than approximately 74$\%$ of the extremal charge.
For positive $\gamma$ we observe that a singularity with diverging curvature invariants 
forms outside the horizon when approaching extremality. 

\end{abstract}
 
%%%%%%%%%%%%%%%%%%%%%%%%%%%%%%%%%%%%%%%%%%%%%%%%%%%%%%%%%%%%%%%%%%%%%%%%%%%%%%
\section{Introduction}
Black holes are a {\it a priori} theoretical prediction of the best theory of gravity that we have to this day, General
Relativity \cite{EinsteinGR}. Recent direct detections of gravitational waves (GWs) \cite{Abbott}
seem to provide mounting evidence that these compact objects -- indeed -- exist in the universe. 
In order to extract data from the detected GW signals, it is very important to understand the 
processes that led to their emission. As such, exact solutions to GR are very important. And although GR is highly non-linear, solutions of this type do exist and are well understood.
Next to the spherically symmetric solutions to the (electro)vacuum Einstein equation \cite{schwarzschild,reissner_nordstrom} -- which are necessarily static \cite{birkhoff,israel} -- stationarily rotating 
(electro)vacuum solutions exist in the form of the Kerr(-Newman) solutions \cite{kerr}, which
are necessarily axisymmetric \cite{Hawking1}. Interestingly, these black hole solutions
are described uniquely by a very small amount of parameters that are subject to a Gauss
law~: mass $M$, charge $Q$ and angular momentum $J$ \cite{carter,robinson,heusler,MTW}.
This fact was summarized in the statement that {\it black holes have no hair}.

The question then arises what happens if next to electromagnetic fields additional matter fields are present. When considering static scalar fields, a number of no-hair theorems
have been proven for asymptotically flat black hole space-times in standard GR. Black holes with regular event horizon can neither support static, massless scalar fields \cite{chase,Bekenstein:1972ny} nor massive scalar fields \cite{Bekenstein:1972ny,teitelboim} nor scalar fields with self-interaction potential and non-negative energy-density \cite{bekenstein:1992}. Interestingly, in \cite{bekenstein:1992} it was also demonstrated that the theorem can be extended to the Brans-Dicke scalar-gravity model. In this latter model, a real scalar field
is non-minimally coupled to gravity by replacing the Einstein-Hilbert term $R/G$ by
$\phi R$, where $R$ is the Ricci scalar \cite{brans_dicke}. The scalar field hence plays the r\^ole of a varying Newton's constant $\phi=G^{-1}$. Brans and Dicke introduced this coupling in order to take Mach's principle into account.

In recent times, scalar-tensor gravity models have become popular again. One reason being the application to cosmology where a scalar field is believed to have driven the very rapid expansion of the universe
very shortly after the Planck era, which is typically referred to as  ``inflation". 
Models that have been discussed extensively in this direction are the so-called Horndeski scalar-tensor gravity models \cite{horndeski},
which constitute all possible scalar-tensor gravity models that lead to second order equations of motion \cite{Deffayet:2013lga,Charmousis:2014mia}. These models are, however, also interesting with view
to the above mentioned no-hair theorems. In contrast to minimally coupled scalars and Brans-Dicke theory,
Horndeski models allow static, asymptotically flat black holes that carry scalar hair \cite{Sotiriou:2014pfa, Babichev:2017ab}. In \cite{Sotiriou:2014pfa} a concrete example was provided using a model that next to the Einstein-Hilbert action contains a scalar-tensor coupling of the form $\phi {\cal G}$, where ${\cal G}$ is the Gauss-Bonnet term. This model possesses a shift symmetry for the
scalar field of the form $\phi \rightarrow \phi + c$, where $c$ is a constant, which leads to
an associated conserved Noether current. 

In the following, models with non-minimal coupling between the scalar field and the metric (as well as other fields) have been discussed. Typically, so-called ``scalarization''
of black holes appears in models that contain non-minimal coupling terms of the form
$f(\phi){\cal I}(g_{\mu\nu};\Sigma)$, where $f(\phi)$ is a function of the scalar field and ${\cal I}$ depends on the metric $g_{\mu\nu}$ and/or other fields $\Sigma$ and  acts as a ``source term'' in the scalar field equation. The first example of this type was given in
a model of a conformally coupled scalar field with interaction term of the form $1/6\phi^2R$ \cite{BBMB}, where $R$ is the Ricci scalar. Recently, the scalarization of static, uncharged black holes with ${\cal I}={\cal G}$ have been
discussed -- for  $f(\phi)=\phi^2$ \cite{Silva:2017uqg} and for different other forms of $f(\phi)$ with
a single tem in $f(\phi)$ \cite{Doneva:2017bvd, Antoniou:2017acq, Antoniou:2017hxj} as well
as a combination of different powers of $\phi$ \cite{Minamitsuji:2018xde}. In all case, the scalarization
appears only for sufficiently large coupling between the scalar field and the GB term. 
In \cite{Brihaye:2018grv} a model combining the original shift symmetric scalar field and a quadratic scalar field coupled to the GB term has been studied bridging between shift symmetry and spontaneous scalarization.  The stability of scalarized, static black holes has also been discussed recently \cite{Silva:2018qhn}. 
The models can be extended to include charge of the black hole. This has been achieved in the conformally coupled scalar field case \cite{BBMB} as well as for a complex scalar field and $f(\phi)$ different from $\phi^2$ \cite{Doneva:2018rou}.
In fact, electromagnetic fields can source the scalar field when non-minimally coupled. This was 
demonstrated for $f(\phi)=\exp(-\alpha\phi^2)$ and ${\cal I}=F_{\mu\nu} F^{\mu\nu}$, where
$F_{\mu\nu}$ is the electromagnetic field strength tensor, in \cite{Herdeiro:2018wub}. Moreover,
models containing higher order terms in $F_{\mu\nu}$ can also lead to scalarization \cite{Herdeiro:2019yjy}. 

In this paper, we are investigating the model discussed in \cite{Doneva:2018rou}, but are mainly interested in the scalarization of near-extremal black holes and show that new features appear.
The RN -- in contrast to the Schwarzschild -- black hole possesses an extremal limit at which the Hawking temperature of the black hole tends
to zero. In this limit, the near-horizon geometry of the RN  is given  by a product of a 2-dimensional
Anti-de Sitter geometry $AdS_2$ and a 2-dimensional sphere $S^2$ with the additional property that the curvature radius of the $AdS_2$ and the radius of the sphere are both equal to the mass of the solution (see e.g. \cite{Carroll} for a detailed
discussion on the geometry of extremal RN solutions). The fact, that an AdS factor appears makes it possible
to associate a dual Conformal Field Theory (CFT) to it via the AdS/CFT correspondence \cite{adscft} and hence compute the
black hole entropy via the dual CFT \cite{Strominger}. 

The RN solution is often considered a simpler ``toy model'' for the Kerr metric \cite{relation_RN_Kerr} as the latter possesses
also an extremal limit and a similar causal structure of the space-time. 
We will follow this point view here and will study the scalarization of charged, static
and spherically symmetric black hole solutions also with the motivation to learn something about the scalarization of stationary solutions with angular momentum.

Our paper is organized as follows~: in Section 2, we will introduce the model and Ansatz.
In Section 3, we will discuss the scalarization of the RN solution, solving the scalar field equation in the background of this black hole solution. Section 4 contains our results on the full back-reacted problem.
We summarize and conclude in Section 5.

\section{The model}
The model we are studying in this paper is a scalar-tensor gravity model that contains a non-minimal coupling between a real scalar field and the Gauss-Bonnet term.  
This model reads~:
\begin{equation}
\label{action}
S =  \int  {\rm d}^4x  \sqrt{-g} \left[\frac{R}{2} + \gamma\phi^2{\cal G}  -  
D_{\mu} \phi  D^{\mu} \phi  -\frac{1}{4}F_{\mu\nu} F^{\mu\nu} \right]   \ ,
\end{equation}
where the Gauss-Bonnet term ${\cal G}$ is given by
\begin{equation}
 {\cal G} = R_{\mu \nu \rho \sigma} R^{\mu \nu \rho \sigma} - 4 R_{\mu \nu}R^{\mu \nu} + R^2  \ ,
\end{equation}
and units are chosen such that $8\pi G\equiv 1$. $D_{\mu}$ denotes the gravitational
covariant derivative, and we keep the covariant notation for clarity, remembering
that $D_{\mu}\phi=\partial_{\mu}\phi$. 

Variation with respect to the scalar field, U(1) gauge field and metric, respectively, leads to the following equations of motion~:
\begin{equation}
\label{eq:scalar}
\square\phi+2\gamma\phi {\cal G}=0   \ ,
\end{equation}
\begin{equation}
\label{eq:em}
\frac{1}{\sqrt{-g}}\partial_{\mu} \left(\sqrt{-g} F^{\mu\nu}\right)=0   \ ,
\end{equation}
\begin{equation}
\label{eq:gravity}
G_{\mu\nu}= D_{\mu}\phi D_{\nu}\phi - \frac{1}{2} g_{\mu\nu} D_{\alpha}\phi D^{\alpha}\phi  - \gamma \left(g_{\mu\rho} g_{\nu\sigma} + g_{\nu\rho} g_{\mu\sigma}\right) \epsilon^{\rho\alpha\beta\gamma} \epsilon^{\delta\sigma\zeta\chi} R_{\beta\gamma\zeta\chi} D_{\alpha} D_{\delta} \left(\phi^2\right)   \ .
\end{equation}

The above model has been discussed for vanishing electromagnetic field  for the first time in \cite{Silva:2017uqg}.  The black hole solutions of this model undergo a {\it spontanenous scalarization} for $\gamma$ sufficiently large and positive. 
In this paper, we are interested in the scalarization of a charged, spherically symmetric black hole. For the metric and the scalar field, we choose the following Ansatz~:
\begin{equation}
\label{ansatz}
ds^2=-N(r) \sigma(r)^2 dt^2 + \frac{1}{N(r)} dr^2  + r^2 \left(d\theta^2 + \sin^2 \theta d \varphi^2\right)  \ \ , \ \  \phi=\phi(r)  \  \  , 
\end{equation}
while the U(1) gauge field is~:
\begin{equation}
A_{\mu}dx^{\mu} = V(r) dt  \ .
\end{equation}
The black holes will therefore be electrically charged. For $\phi(r)\equiv 0$, the model has
a spherically symmetric, static solution~: the Reissner-Nordstr\"om solution \cite{reissner_nordstrom} which is uniquely determined by its ADM mass and (electric and/or magnetic) charge.

\section{Scalarization of a Reissner-Nordstr\"om black hole}
\label{subsec:background}
Here we will be interested in discussing the effect that the electric charge of a Reissner-Nordstr\"om (RN) solution can have on the process of spontaneous scalarization in our model. For that, we have first solved the scalar field equation in the background of the RN solution, which using our conventions reads~:
\begin{equation}
ds^2 = -N dt^2 + \frac{1}{N} dr^2 + r^2\left(d\theta^2 + \sin^2\theta d\varphi^2\right) \ \ , \ \ 
N(r)=1-\frac{2M}{r} + \frac{Q^2}{2r^2} \ \  , \ \    V(r)=\frac{Q}{r_h} - \frac{Q}{r}  \  ,
\end{equation}
where we haved fixed $V(r_h)=0$.
$M$ denotes the ADM mass and $Q$ the electric charge of the solution. The event horizon of
this solution is at $r_h=M+\sqrt{M^2 - Q^2/2}$ with extremal limit at $r_h=M=Q/\sqrt{2}$. In the following, we will
fix $r_h=1$, which determines the mass $M$ in terms of the electric charge $Q$ via the relation
$M=(Q^2+2)/4$. The extremal limit is then at $r_h=M=1$, $Q=\sqrt{2}$.

The scalar field equation (\ref{eq:scalar}) in the background of the RN solution reads~:
\begin{equation}
\label{eq:scalar_background}
\frac{1}{r^2} \left(r^2 N \phi'\right)' = -\gamma_2 \phi {\cal G}_{\rm RN}
\end{equation}
with the Gauss-Bonnet term of the RN solution given by~:
\begin{equation}
\label{gb_rn}
{\cal G}_{\rm RN}=\frac{12}{r^6} + \frac{12(r-2) Q^2}{r^7} + \frac{(3r^2 - 12r +10)Q^4}{r^8} \ .
\end{equation}
The prime now and in the following denotes the derivative with respect to $r$.

To solve (\ref{eq:scalar_background}), we need to fix the appropriate boundary conditions. These are~:
\begin{equation}
\label{eq:bc_RN}
\frac{\phi'(r_h)}{\phi(r_h)}=2\gamma_2 \frac{12-12Q^2+Q^4}{2-Q^2} \ \ \ , \ \ \ 
\phi(r\rightarrow \infty) = \frac{Q_{\rm s}}{r} + {\rm O}(r^{-2}) \ .
\end{equation}
The condition on the horizon $r=r_h$ is related to the requirement of regularity of the scalar field, while the condition at infinity determines the scalar charge $Q_{\rm s}$ of the solution. We also
fix $\phi(r_h)=1$. 

We have solved (\ref{eq:scalar_background}) numerically using a collocation solver for ordinary differential equations \cite{colsys}. Note that for a fixed $Q$, the boundary condition fixes the derivative of the
scalar field function at the horizon and that scalarized black hole exist only for a specific value of $\gamma(Q)$. 

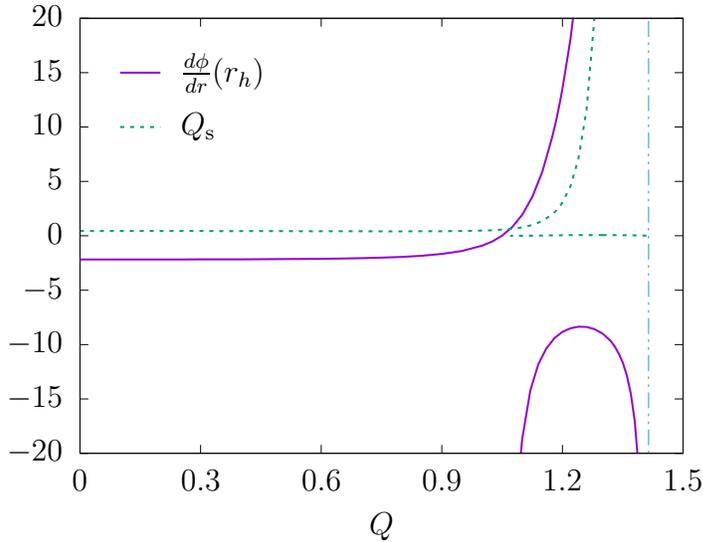
\begin{figure}[ht!]
\begin{center}
\input{phip_qs_q}
\caption{We show the value of the derivative of the scalar field at the horizon, $\frac{d\phi}{dr}(r_h)$, (solid purple) as well as the scalar charge of the solution, $Q_{\rm s}$, (dashed green) as function of the electric charge $Q$ of the RN black hole. Note that the vertical line at $Q=\sqrt{2}$ (dotted-dashed blue)
corresponds to the extremal charge of a RN black hole with $r_h=1$. }
\label{fig:phip_qs_q}
\end{center}
\end{figure}

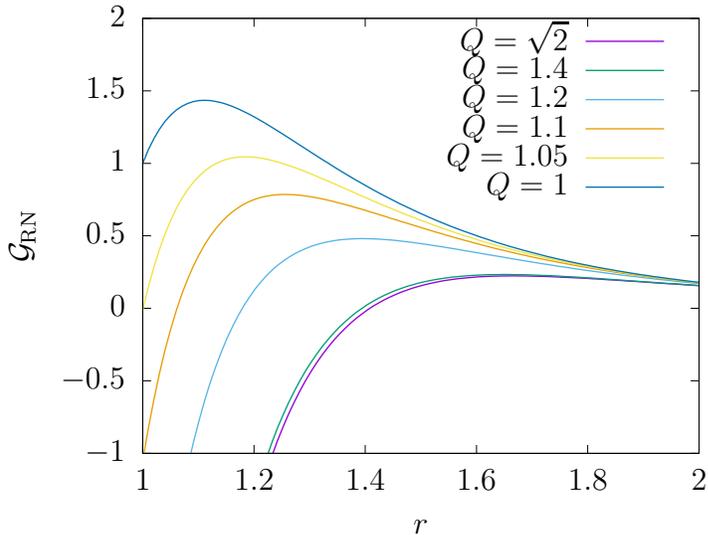
\begin{figure}[ht!]
\begin{center}
\input{GB_RN}
\caption{We show the Gauss-Bonnet term ${\cal G}_{\rm RN}$ of the RN solution close to
the horizon $r_h=1$ for different
value of the electric charge $Q$. Note that $Q=\sqrt{2}$ corresponds to the extremal RN solution. }
\label{fig:gb_rn}
\end{center}
\end{figure}

\begin{figure}[ht!]
\begin{center}
\input{profile_q14}
\caption{We show the profile of the scalar field function $\phi(r)$ for $Q=1.4$ (i.e. a value of the charge close to the extremal limit $Q=\sqrt{2}$) on the two different
branches~: for $\gamma=1.954$ (solid purple) and for $\gamma=-0.082$ (dotted-dashed green), respectively.  }
\label{fig:profile_q14}
\end{center}
\end{figure}
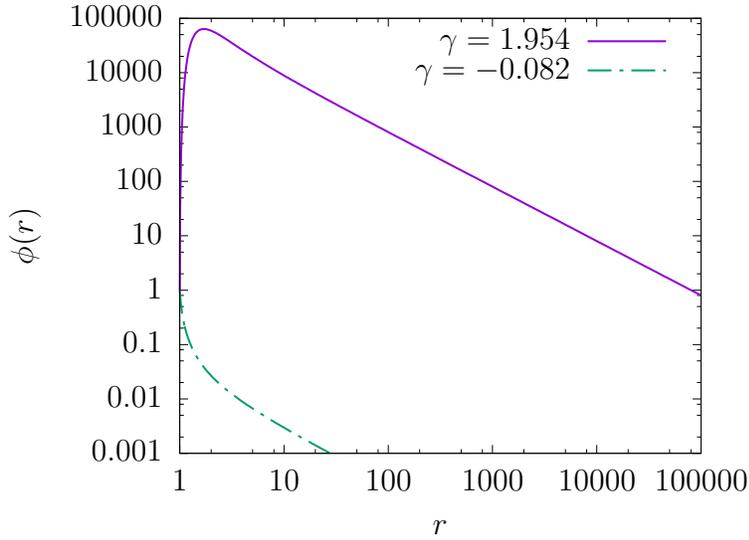

Our results are shown in Fig.~\ref{fig:phip_qs_q}, where we give the
derivative of the scalar field at the horizon $\frac{d\phi}{dr}(r_h)\equiv \phi'(r_h)$ as well as the scalar charge $Q_{\rm s}$ of the solution as function of the electric charge $Q$.
For $Q=0$, we recover the result of \cite{Silva:2017uqg}. Increasing $Q$ from zero, a branch of scalarized black holes exists for positive values of $\gamma$ with scalar charge $Q_{\rm s}$ 
varying only marginally. However, 
the boundary condition at $r=r_h$ indicates that
for $Q >\tilde{Q} = \sqrt{2\left(3-\sqrt{6}\right)}\approx 1.05$ the derivative of $\phi$ at the horizon, $\phi'(r_h)$, 
changes sign. This is, in fact, connected to the change of the
sign of the GB term ${\cal G}_{\rm RN}$ at (and close to) the horizon $r_h=1$ -- see (\ref{gb_rn}). To demonstrate this, we show the Gauss-Bonnet term of the RN solution close to the horizon and for different values of $Q$ in Fig.~\ref{fig:gb_rn}. While for $Q < \tilde{Q}$ the Gauss-Bonnet term is positive everywhere outside the horizon, this is no longer true for $Q > \tilde{Q}$.  When increasing the charge $Q$,
the interval in $r$ for which ${\cal G}_{\rm RN}$ is negative
increases. This is not possible in the case of an uncharged black hole
and is fundamentally connected to the $AdS_2\times S^2$ near-horizon geometry of nearly extremally charged RN black holes
(see Appendix for more details).
Now remembering that it is a tachyonic instability in
which the term $\gamma{\cal G}_{\rm RN}$ acts as an  ``effective mass'' of the scalar field, the requirement that $\gamma > 0$ for the instability to appear is certainly no longer valid here. 
In fact, this explains the existence of a second branch of solutions of scalarized black holes that exists for $Q > \tilde{Q}$ and $\gamma < 0$. This is shown in Fig.~\ref{fig:phip_qs_q}. This branch possesses scalar charge $Q_{\rm s}$ small and positive. 
In order to show the difference between the solutions for a fixed value of $Q$, we give the profile of the scalar field function $\phi(r)$ for $Q=1.4$ on the two branches in Fig.~\ref{fig:profile_q14}.
Close to the extremal limit of $Q=\sqrt{2}$, the derivative
of the scalar field at the horizon is already very large. This is to 
be expected from the boundary condition at $r_h$, see (\ref{eq:bc_RN}), which indicates that $\phi'(r_h)\rightarrow \pm \infty$ for $Q\rightarrow \sqrt{2}$. This is also shown in Fig.~\ref{fig:phip_qs_q}, where  we demonstrate that
for $Q\rightarrow \sqrt{2}$, the value of the derivative of the
scalar field function at the horizon, $\phi'(r_h)\rightarrow \pm \infty$ for $\gamma \gtrless 0$. 

The profiles of the scalar field function in Fig. \ref{fig:profile_q14} also show that
the scalar field is small on the negative-$\gamma$ branch, while it becomes very large on the positive-$\gamma$ branch. In fact, for positive $\gamma$, $\phi(r)$ possesses a maximum
outside the horizon $r_h$ and extends to very large values of $r$ ($\phi(r)$ tends to zero
at $r\approx 10^6$). This is very different for negative values of $\gamma$, where the scalar field is very small and tends to zero quickly. The maximum of the scalar field in this
latter case is always at the horizon. Since the space-time is not dynamical, it is only the
scalar field behaviour that leads to the critical behaviour. When taking the backreaction of the space-time into account (see Section \ref{section:backreaction}), we find that the behaviour of the scalar field strongly influences the space-time and leads to the appearance of a metric singularity.

\section{Including backreaction} 
\label{section:backreaction}
In order to solve the full set of coupled non-linear differential equations numerically, we have to employ the appropriate boundary conditions. At the regular horizon $r_h$ these are
\begin{equation}
N(r_h)=0 \ \ , \ \  A (\phi')^2 + B \phi + C = 0 \ ,
\end{equation}
where $A$, $B$, $C$ are given as follows
\begin{eqnarray}
A &=&  \left .  \left[8\gamma \sigma^2 \phi \left(2 \sigma^2 r_h^2 - 64 \gamma^2 \phi^2 (V')^2  - (V')^2 r_h^4\right)\right]\right\vert_{r=r_h}  \ , \nonumber \\
B & =& r_h \left. \left[(V')^2 \sigma^2 \left(-r_h^4 - 64\gamma^2 \phi^2 \right) + 2 \sigma^4 r_h^3
-32 r_h^2 \gamma^2 \phi^2 (V')^4 \right]\right\vert_{r=r_h}  \ , \nonumber \\
C & =& \left. \left[2\gamma\phi \left(12 \sigma^4 - 12 \sigma^2 (V')^2 r_h^2 + (V')^4 r_h^4\right) \right]\right\vert_{r=r_h}  \ ,
\end{eqnarray}
such that the boundary condition for $\phi'$ at the horizon $r_h$ is
\begin{equation}
\phi'\vert_{r=r_h} = \frac{-B \pm \sqrt{B^2 - 4 AC}}{2A}  \ .
\end{equation}
The existence of scalarized black holes is hence limited by the requirement that $B^2 - 4 AC \geq 0$.
In the following it will be useful to write this condition in terms of two factors as $B^2-4C = \Delta_1^2 \cdot \Delta_2$ with
\begin{eqnarray}
\label{eq:delta12}
\Delta_1 &=& \left.\left[2\sigma^2 - V'^2 r_h^2\right]\right\vert_{r=r_h} \ , \  \nonumber \\
\Delta_2 &=& \left. \left[\sigma^4 r_h^2 \left(r_h^4 - 384 \gamma^2 \phi^2\right)  + 128 V'^2 \sigma^2 \phi^2 \gamma^2 \left(r_h^4 + 96 \gamma^2 \phi^2\right) + 1024 \gamma^4 \phi^4 r_h^2  V'^4\right)\right\vert_{r=r_h}
\end{eqnarray}

In order for the solutions to be asymptotically flat and have finite energy, we require
\begin{equation}
\sigma(r\rightarrow \infty)=1 \ \ , \ \    V(r\rightarrow \infty)=-\frac{Q}{r} + {\rm O}(r^{-2}) \ \  ,  \ \ 
\phi(r\rightarrow \infty) = \frac{Q_{\rm s}}{r} + {\rm O}(r^{-2}) 
\end{equation}

\begin{figure}[ht!]
\begin{center}
\input{domain_q_ga}
\caption{We show the domain of existence of scalarized, charged black holes in the $\gamma$-$Q$-plane. 
Solutions exist for values $\gamma\in [\gamma_{0}:\gamma_{\rm cr}]$, where $\gamma_{0}$ (solid purple) corresponds to the value of $\gamma$ for which $\phi(r) << 1$ or -- equivalently -- to the value of
$\gamma$ for which solutions of the scalar field equation in the background of the RN solution exist (see Section \ref{subsec:background}). $\gamma_{cr}$ (dotted-dashed green) corresponds to the critical value of $\gamma$ to where
scalarized black holes exist with the maximal possible value of $\phi(r_h)$. The division of the domain into the two different approaches to criticality is indicated by a vertical line (dashed orange) at $Q=1.05$, while the extremal value of the charge is indicated by a vertical line (dashed blue)
at $Q=\sqrt{2}\approx 1.414$. }
\label{fig:domain_q_ga}
\end{center}
\end{figure}
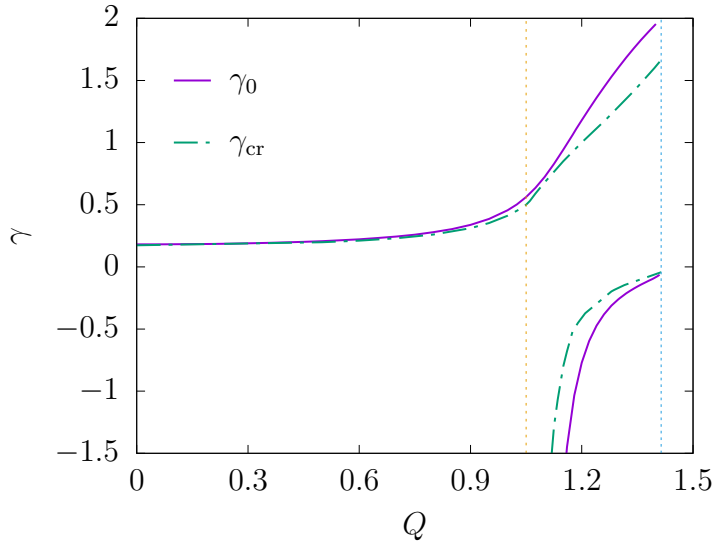

In the following, we have fixed the horizon radius to $r_h=1$ without loss of generality and have constructed black hole solutions numerically for different values of $Q$ and $\gamma$. 
We find that for fixed values of $Q$ scalarized black holes exist only in an interval 
$\gamma\in [\gamma_0:\gamma_{\rm cr}]$. This is shown in Fig.~\ref{fig:domain_q_ga}. For positive values of $\gamma$, we find that
$\gamma_{0}$ corresponds to the value of $\gamma$ for which $\phi(r_h)\rightarrow 0$. This value is equivalent to the value of $\gamma$ for which the scalar field
solutions in the background of the RN black hole exist (see Section \ref{subsec:background}).
For a fixed $Q$, we have then decreased the value of $\gamma$  (which is equivalent 
to increasing the value of $\phi(r_h)$) down to a value of $\gamma_{\rm cr}$, where
the solution with the largest possible value of $\phi(r_h)$ exists. For $Q \leq \tilde{Q}=1.05$, the solutions stop because $\Delta_2\rightarrow 0$. Moreover, as is obvious from Fig.~\ref{fig:domain_q_ga}, $\gamma_{0}$ and $\gamma_{\rm cr}$ are very close to each other. We give the numerical values for some charges in Table \ref{table:gamma_q}.

\begin{table}
\begin{center}
  \begin{tabular}{ | r | r | r | r | r |}
    \hline
    $Q$ & $\gamma^{(+)}_{0}$ & $\gamma^{(+)}_{\rm cr}$ & $\gamma^{(-)}_{\rm 0}$ & $\gamma^{(-)}_{\rm cr}$ \\   \hline
    $0$ & $ 0.181$  & $0.173$ & $-$ & $-$ \\ 
    $0.5$ & $0.207$   & $0.198$ & $-$ & $-$    \\ 
     $1.0$ & $0.455$   & $0.414$ &  $-$ & $-$        \\ 
      $1.1$ & $0.724$   & $ 0.680$ & $-6.980$  & $-2.100$  \\ 
        $1.2$ & $1.180$   & $1.015$ & $-0.770$ & $-0.400$  \\ 
     $1.4$ & $1.954$   & $1.550$ & $-0.082$ & $-0.067$  \\
    \hline
  \end{tabular}
  \caption{We give the values of $\gamma$ where $\phi(r_h)\rightarrow 0$, $\gamma_{0}$, and the
  critical value of $\gamma$, $\gamma_{\rm cr}$ to where scalarized black holes exist with
  the maximal possible value of $\phi(r_h)$. The upper index $(+)$ (respectively $(-)$) indicates the branch with positive (negative) values of $\gamma$. When no negative $\gamma$ is given, only the positive $\gamma$ branch exists.}
  \label{table:gamma_q}
  \end{center}
\end{table}

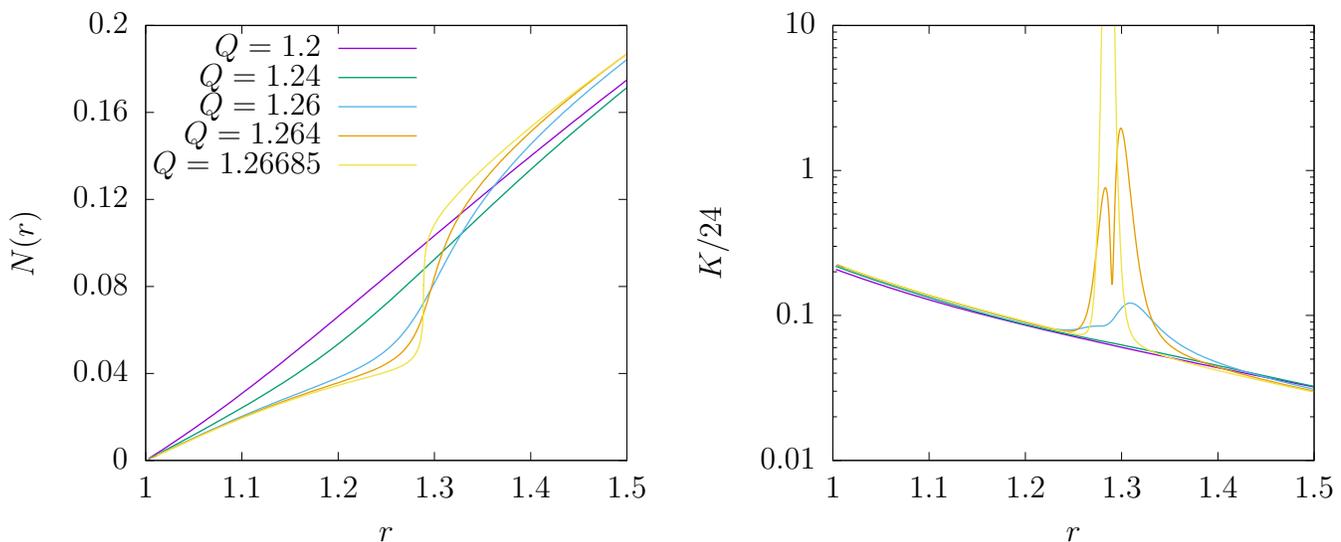
\begin{figure}[ht!]
\begin{center}
\input{profile_N_phi0_001.tex}
\input{profile_kretsch_phi0_001.tex}
\caption{We show the profiles of the metric functions $N(r)$ (left) and the Kretschmann scalar $K=R_{\mu\nu\sigma\rho} R^{\mu\nu\sigma\rho}$ (right, same colour coding as left) close to the horizon
$r_h=1$ for $\phi(r_h)=0.01$. }
\label{fig:criticality}
\end{center}
\end{figure}

\begin{figure}[ht!]
\begin{center}
\input{profile_vp_phi0_001.tex}
\input{profile_phi_phi0_001.tex}
\caption{We show the profiles of the electric field $E(r) = -\frac{dV}{dr}$ (left) and of the scalar field function $\phi(r)$ (right, same colour coding as left) close to the horizon
$r_h=1$ for $\phi(r_h)=0.01$. }
\label{fig:criticality2}
\end{center}
\end{figure}
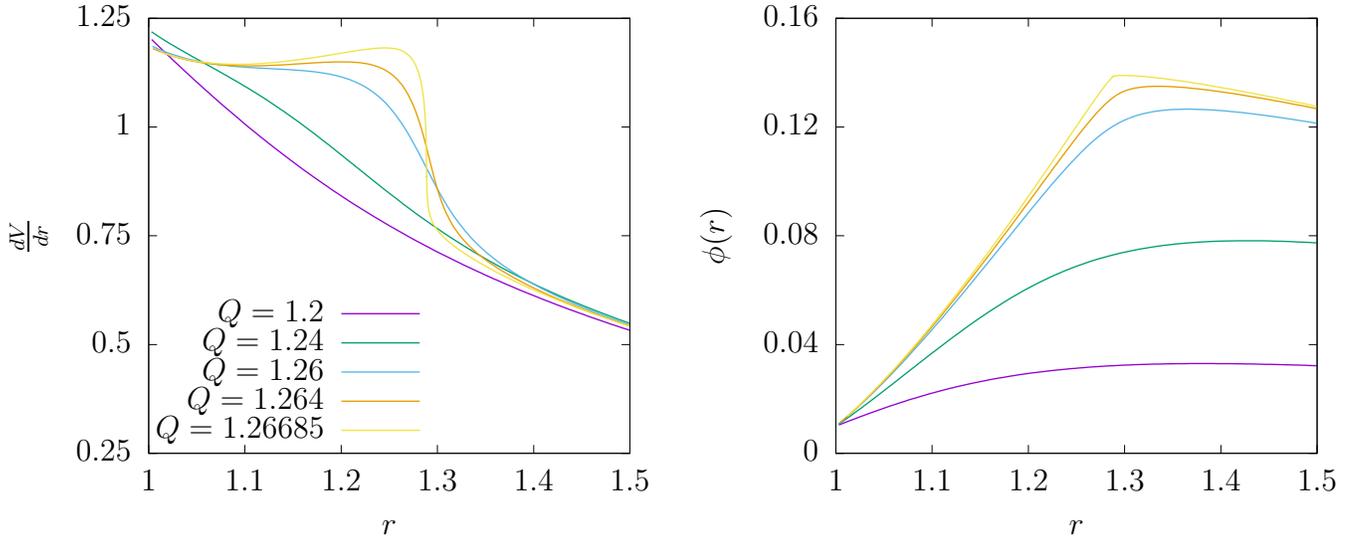

For $Q > \tilde{Q}$, the approach to criticality is very different. When decreasing $\gamma$ (or equivalently increasing $Q$ or $\phi(r_h)$), we find that 
a singularity in the metric curvature starts to form. We show this phenomenon for fixed value of
$\phi(r_h)=0.01$ and increasing $Q$ in Fig.\ref{fig:criticality}  and Fig.\ref{fig:criticality2}.
The profile of the metric function $N(r)$ as given in Fig.\ref{fig:criticality} demonstrates that
at a specific radius outside the horizon, $r=r_{\infty}$, the derivative of $N(r)$ becomes infinite. This corresponds to a strong increase in the metric curvature -- see the behaviour of the Kretschmann scalar $K=R_{\mu\nu\rho\sigma} R^{\mu\nu\rho\sigma}$ given in Fig.\ref{fig:criticality}. At the same time, the electric field of the solution $E(r)=-\frac{dV}{dr}$ (see Fig.\ref{fig:criticality2}) develops an infinite derivative at $r_{\infty}$, while the
scalar field itself becomes non-differentiable there (see Fig.\ref{fig:criticality2}). 
For the solution shown in Fig.\ref{fig:criticality} and Fig.\ref{fig:criticality2}, we find that
$r_{\infty}\approx 1.29$ and $\gamma_{\rm cr}\approx 1.19$. For increasing (decreasing)
$\gamma_{\rm cr}$, we find that $r_{\rm \infty}$ increases (decreases), e.g. we find that
for $\gamma_{\rm cr}\approx 1.40$ (corresponding to $Q=1.342$) $r_{\rm \infty}\approx 1.40$, while for
$\gamma_{\rm cr}\approx 0.68$ (which corresponds to $Q=1.1$) $r_{\rm \infty}\approx 1.09$. All our results indicate that $r_{\infty}\rightarrow r_h=1$ for $Q\rightarrow \tilde{Q}=1.05$. 
The reason for this behaviour becomes clear when investigating the behaviour of the GB term close and on the horizon. For $Q > \tilde{Q}$, the GB term is negative on the horizon, but possesses
a zero somewhere outside the horizon at $r=r_0$ such that for $r > r_0$ the GB term is positive.
This value of $r_0$ separates the interval of $r$ into two intervals~: a) $r\in [r_h:r_0]$ for which the GB term 
acts as a regular (position dependent) ``mass term'' for the scalar field (see also the Appendix for
more details) and b) $r > r_0$ for which the tachyonic instability persists. Not surprisingly, the
behaviour at the intersection between the two intervals becomes discontinuous. In fact,
we have numerically confirmed that the value $r_{\infty}\approx r_0$. 

Similar to the case of the scalar field in the background of the RN solution, we find that scalarized
black holes exist for negative values of $\gamma$ when $Q$ is sufficiently large. 
These solutions exist for $Q > \tilde{Q}=1.05$. We find a similar pattern as is the case of
positive $\gamma$ with $\phi(r_h)\rightarrow 0$ for  $\gamma_{0}$ and $\phi(r_h)$ having its
maximal possible value for $\gamma\rightarrow\gamma_{\rm cr}$. 
At $\gamma_{\rm cr}$ we find that the solutions stop because the metric curvature
diverges, indicated by a divergence of the Kretschmann scalar $K$, but at the horizon
$r_h$ itself, and not, as in the case for positive $\gamma$ at a point outside the horizon.
The values of $\gamma_0$ and $\gamma_{\rm cr}$ are given for some values of $Q$ in
Table \ref{table:gamma_q}. Note that in this case,  the negative GB term and negative value of
$\gamma$ leads to the appearance of the tachyonic instability on and close to the horizon.
Since the GB becomes positive and together with the negative value of $\gamma$ becomes
a ``standard mass term'' on and outside the horizon, a metric singularity outside the horizon does not
appear in this case.

\section{Conclusions and Outlook}
In this paper, we have studied the scalarization of charged, static, spherically symmetric black holes
in a quadratic Einstein-scalar-Gauss-Bonnet model. We observe that for sufficiently large values of the electric charge $Q$ (at approximately $74\%$ of extremality), new phenomena appear in contrast to
uncharged and mildly charged black holes. We observe the formation of a curvature singularity
outside the event horizon as well as the existence of new solutions for negative values of the
coupling constant. The reason for this is that the Gauss-Bonnet term that sources the scalar field
becomes negative close and on the horizon. Depending on the choice of coupling, the source
terms hence either acts as a standard, distance dependent ``mass term'' or as a term that
drives the tachyonic instability and hence the scalarization of the black hole. 

The system possesses solutions with radial excitations of the scalar field as well as angular excitations related to higher multipoles. We will report on these solutions in the future. 
Another interesting extension of our results would be to add a term of the form $\phi^2 R$ as well as
a fourth-order self-interaction of the scalar field to our model. Without the GB term this model
describes Higgs inflation at constant, non-running coupling constants \cite{Bezrukov1}. An interesting
question is whether primordial black holes can form during inflation. Black hole production
in Higgs inflation has been discussed recently \cite{Ezquiaga:2017fvi}.
Clearly, without the GB term, the black holes would be standard RN black holes as
a quick investigation of the ``tachyonic instability'' argument shows. 
However, as we have shown in this paper, scalarized black holes with small (negative) value of
$\gamma$ can be constructed and it would be interesting to see how these solutions are influenced
by the presence of the aforementioned terms. This is currently under investigation. 

\vspace{1cm}

{\bf Acknowledgements} 
 BH would like to thank FAPESP for financial support under
grant number {\it 2016/12605-2} and CNPq for financial support under
{\it Bolsa de Produtividade Grant 304100/2015-3}.  

\clearpage

%%%%%%%%%%%%%%%%%%%%%%%%%%%%%%%%%%%%%%%%%%%%%%%%%%%%%%%%%%%%

%%%%%%%%%%%%%%%%%%%%%%%%%%%%%%%%%%%%%%%%%%%%%%%%%%%%%%%%%%%%

\section{Appendix: A tachyonic instability for extremal RN black holes}
As noted in \cite{Silva:2017uqg}, the spontaneous scalarization of a black hole in a model containing
a non-minimal coupling between the scalar field and the Gauss-Bonnet term is related to the appearance of a tachyonic instability of the scalar field. The reasoning in this latter paper is as follows~: assume the background space-time to be fixed and write the scalar field perturbation $\delta \phi$ 
as $\delta\phi =\frac{f(r,t)}{r}Y_{\ell m}(\theta,\varphi)$ 
with the $Y_{\ell m}$ being the spherical harmonics. 

As background we choose the extremally charged RN solution which has $r_+=M=Q/\sqrt{2}$ and hence can be written as
\begin{equation}
ds^2 = - \left(1 - \frac{M}{r}\right)^2 dt^2 +  \left(1 - \frac{M}{r}\right)^{-2} dr^2 + r^2 \left(d\theta^2 + \sin^2 \theta d \varphi^2\right)
\end{equation}
which taking the near-horizon limit $t\rightarrow \tilde{t}/\epsilon$, $r\rightarrow M + \epsilon \rho$, $\epsilon\rightarrow 0$ and then substituting $\tilde{t}= M^2 \tau$ becomes
\begin{equation}
\label{ads2s2}
ds^2 - M^2\left(\rho^2  d\tau^2 + \frac{1}{\rho^2} d\rho^2\right) + M^2\left(d\theta^2 + \sin^2 \theta d \varphi^2\right)  \ .
\end{equation}
This is the well-known result that extremal black hole solutions possess an $AdS_2\times S^2$ near horizon geometry, in this case with $AdS$ radius $L$ and radius $R_{S^2}$ of the $S^2$ fulfiling
$L=R_{S^2}=M=r_+=Q/\sqrt{2}$.

Introducing the coordinate
$\rho_{*}=\rho^{-1}$ and inserting the Gauss-Bonnet term of the metric (\ref{ads2s2})
${\cal G}_{\rm exRN}=-8/M^4$ the equation for the scalar field perturbation $\delta\phi$ reads~:
\begin{equation}
\label{eq:exRN}
-\frac{\partial^2 f}{\partial \tau^2} + \frac{\partial^2 f}{\partial \rho_{*}^2}= 
\left(\frac{\ell(\ell+1)}{\rho_{*}^2} Y_{\ell m}  + \frac{16\gamma}{M^4}       \right) f \ .
\end{equation}
For $\gamma > 0$, the GB term now acts like a ``standard'' mass term with
static, $\ell=0$ solutions of the form $f(\rho)= A\exp\left(-\kappa/\rho\right) + B \exp\left(\kappa/\rho\right)$
with $\kappa:=4\sqrt{\gamma}/M^2$ and $A$, $B$ integration constants. 
These solutions are either zero or tend to infinity for $\rho\rightarrow 0$, i.e. at the horizon.

The tachyonic instability as mentioned above appears for $\gamma < 0$. Solutions
to (\ref{eq:exRN}) are then of the form $f(\rho)=a\sin(k/\rho)+b\cos(k/\rho)$, where
$k:=4\sqrt{-\gamma}/M^2$  and $a$, $b$ are integration constants.

 \end{document}

%% file: phip_qs_q.tex
% GNUPLOT: LaTeX picture with Postscript
\begingroup
  \makeatletter
  \providecommand\color[2][]{%
    \GenericError{(gnuplot) \space\space\space\@spaces}{%
      Package color not loaded in conjunction with
      terminal option `colourtext'%
    }{See the gnuplot documentation for explanation.%
    }{Either use 'blacktext' in gnuplot or load the package
      color.sty in LaTeX.}%
    \renewcommand\color[2][]{}%
  }%
  \providecommand\includegraphics[2][]{%
    \GenericError{(gnuplot) \space\space\space\@spaces}{%
      Package graphicx or graphics not loaded%
    }{See the gnuplot documentation for explanation.%
    }{The gnuplot epslatex terminal needs graphicx.sty or graphics.sty.}%
    \renewcommand\includegraphics[2][]{}%
  }%
  \providecommand\rotatebox[2]{#2}%
  \@ifundefined{ifGPcolor}{%
    \newif\ifGPcolor
    \GPcolortrue
  }{}%
  \@ifundefined{ifGPblacktext}{%
    \newif\ifGPblacktext
    \GPblacktextfalse
  }{}%
  % define a \g@addto@macro without @ in the name:
  \let\gplgaddtomacro\g@addto@macro
  % define empty templates for all commands taking text:
  \gdef\gplbacktext{}%
  \gdef\gplfronttext{}%
  \makeatother
  \ifGPblacktext
    % no textcolor at all
    \def\colorrgb#1{}%
    \def\colorgray#1{}%
  \else
    % gray or color?
    \ifGPcolor
      \def\colorrgb#1{\color[rgb]{#1}}%
      \def\colorgray#1{\color[gray]{#1}}%
      \expandafter\def\csname LTw\endcsname{\color{white}}%
      \expandafter\def\csname LTb\endcsname{\color{black}}%
      \expandafter\def\csname LTa\endcsname{\color{black}}%
      \expandafter\def\csname LT0\endcsname{\color[rgb]{1,0,0}}%
      \expandafter\def\csname LT1\endcsname{\color[rgb]{0,1,0}}%
      \expandafter\def\csname LT2\endcsname{\color[rgb]{0,0,1}}%
      \expandafter\def\csname LT3\endcsname{\color[rgb]{1,0,1}}%
      \expandafter\def\csname LT4\endcsname{\color[rgb]{0,1,1}}%
      \expandafter\def\csname LT5\endcsname{\color[rgb]{1,1,0}}%
      \expandafter\def\csname LT6\endcsname{\color[rgb]{0,0,0}}%
      \expandafter\def\csname LT7\endcsname{\color[rgb]{1,0.3,0}}%
      \expandafter\def\csname LT8\endcsname{\color[rgb]{0.5,0.5,0.5}}%
    \else
      % gray
      \def\colorrgb#1{\color{black}}%
      \def\colorgray#1{\color[gray]{#1}}%
      \expandafter\def\csname LTw\endcsname{\color{white}}%
      \expandafter\def\csname LTb\endcsname{\color{black}}%
      \expandafter\def\csname LTa\endcsname{\color{black}}%
      \expandafter\def\csname LT0\endcsname{\color{black}}%
      \expandafter\def\csname LT1\endcsname{\color{black}}%
      \expandafter\def\csname LT2\endcsname{\color{black}}%
      \expandafter\def\csname LT3\endcsname{\color{black}}%
      \expandafter\def\csname LT4\endcsname{\color{black}}%
      \expandafter\def\csname LT5\endcsname{\color{black}}%
      \expandafter\def\csname LT6\endcsname{\color{black}}%
      \expandafter\def\csname LT7\endcsname{\color{black}}%
      \expandafter\def\csname LT8\endcsname{\color{black}}%
    \fi
  \fi
    \setlength{\unitlength}{0.0500bp}%
    \ifx\gptboxheight\undefined%
      \newlength{\gptboxheight}%
      \newlength{\gptboxwidth}%
      \newsavebox{\gptboxtext}%
    \fi%
    \setlength{\fboxrule}{0.5pt}%
    \setlength{\fboxsep}{1pt}%
\begin{picture}(5668.00,4250.00)%
    \gplgaddtomacro\gplbacktext{%
      \csname LTb\endcsname%
      \put(594,704){\makebox(0,0)[r]{\strut{}$-20$}}%
      \put(594,1114){\makebox(0,0)[r]{\strut{}$-15$}}%
      \put(594,1524){\makebox(0,0)[r]{\strut{}$-10$}}%
      \put(594,1934){\makebox(0,0)[r]{\strut{}$-5$}}%
      \put(594,2345){\makebox(0,0)[r]{\strut{}$0$}}%
      \put(594,2755){\makebox(0,0)[r]{\strut{}$5$}}%
      \put(594,3165){\makebox(0,0)[r]{\strut{}$10$}}%
      \put(594,3575){\makebox(0,0)[r]{\strut{}$15$}}%
      \put(594,3985){\makebox(0,0)[r]{\strut{}$20$}}%
      \put(726,484){\makebox(0,0){\strut{}$0$}}%
      \put(1635,484){\makebox(0,0){\strut{}$0.3$}}%
      \put(2544,484){\makebox(0,0){\strut{}$0.6$}}%
      \put(3453,484){\makebox(0,0){\strut{}$0.9$}}%
      \put(4362,484){\makebox(0,0){\strut{}$1.2$}}%
      \put(5271,484){\makebox(0,0){\strut{}$1.5$}}%
      \put(1484,3165){\makebox(0,0)[l]{\strut{}$Q_{\rm s}$}}%
      \put(1484,3575){\makebox(0,0)[l]{\strut{}$\frac{d\phi}{dr}(r_h)$}}%
    }%
    \gplgaddtomacro\gplfronttext{%
      \csname LTb\endcsname%
      \put(2998,154){\makebox(0,0){\strut{}$Q$}}%
    }%
    \gplbacktext
    \put(0,0){\includegraphics{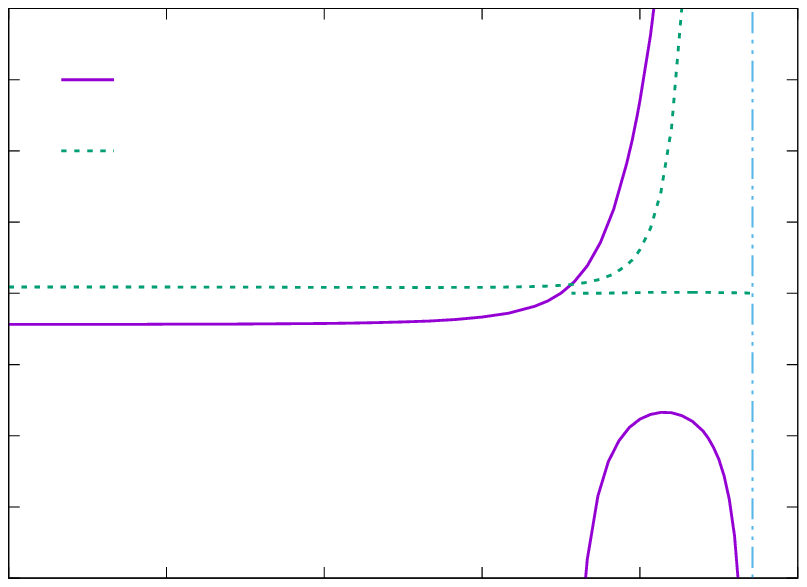}}%
    \gplfronttext
  \end{picture}%
\endgroup

%% file: GB_RN.tex
% GNUPLOT: LaTeX picture with Postscript
\begingroup
  \makeatletter
  \providecommand\color[2][]{%
    \GenericError{(gnuplot) \space\space\space\@spaces}{%
      Package color not loaded in conjunction with
      terminal option `colourtext'%
    }{See the gnuplot documentation for explanation.%
    }{Either use 'blacktext' in gnuplot or load the package
      color.sty in LaTeX.}%
    \renewcommand\color[2][]{}%
  }%
  \providecommand\includegraphics[2][]{%
    \GenericError{(gnuplot) \space\space\space\@spaces}{%
      Package graphicx or graphics not loaded%
    }{See the gnuplot documentation for explanation.%
    }{The gnuplot epslatex terminal needs graphicx.sty or graphics.sty.}%
    \renewcommand\includegraphics[2][]{}%
  }%
  \providecommand\rotatebox[2]{#2}%
  \@ifundefined{ifGPcolor}{%
    \newif\ifGPcolor
    \GPcolortrue
  }{}%
  \@ifundefined{ifGPblacktext}{%
    \newif\ifGPblacktext
    \GPblacktextfalse
  }{}%
  % define a \g@addto@macro without @ in the name:
  \let\gplgaddtomacro\g@addto@macro
  % define empty templates for all commands taking text:
  \gdef\gplbacktext{}%
  \gdef\gplfronttext{}%
  \makeatother
  \ifGPblacktext
    % no textcolor at all
    \def\colorrgb#1{}%
    \def\colorgray#1{}%
  \else
    % gray or color?
    \ifGPcolor
      \def\colorrgb#1{\color[rgb]{#1}}%
      \def\colorgray#1{\color[gray]{#1}}%
      \expandafter\def\csname LTw\endcsname{\color{white}}%
      \expandafter\def\csname LTb\endcsname{\color{black}}%
      \expandafter\def\csname LTa\endcsname{\color{black}}%
      \expandafter\def\csname LT0\endcsname{\color[rgb]{1,0,0}}%
      \expandafter\def\csname LT1\endcsname{\color[rgb]{0,1,0}}%
      \expandafter\def\csname LT2\endcsname{\color[rgb]{0,0,1}}%
      \expandafter\def\csname LT3\endcsname{\color[rgb]{1,0,1}}%
      \expandafter\def\csname LT4\endcsname{\color[rgb]{0,1,1}}%
      \expandafter\def\csname LT5\endcsname{\color[rgb]{1,1,0}}%
      \expandafter\def\csname LT6\endcsname{\color[rgb]{0,0,0}}%
      \expandafter\def\csname LT7\endcsname{\color[rgb]{1,0.3,0}}%
      \expandafter\def\csname LT8\endcsname{\color[rgb]{0.5,0.5,0.5}}%
    \else
      % gray
      \def\colorrgb#1{\color{black}}%
      \def\colorgray#1{\color[gray]{#1}}%
      \expandafter\def\csname LTw\endcsname{\color{white}}%
      \expandafter\def\csname LTb\endcsname{\color{black}}%
      \expandafter\def\csname LTa\endcsname{\color{black}}%
      \expandafter\def\csname LT0\endcsname{\color{black}}%
      \expandafter\def\csname LT1\endcsname{\color{black}}%
      \expandafter\def\csname LT2\endcsname{\color{black}}%
      \expandafter\def\csname LT3\endcsname{\color{black}}%
      \expandafter\def\csname LT4\endcsname{\color{black}}%
      \expandafter\def\csname LT5\endcsname{\color{black}}%
      \expandafter\def\csname LT6\endcsname{\color{black}}%
      \expandafter\def\csname LT7\endcsname{\color{black}}%
      \expandafter\def\csname LT8\endcsname{\color{black}}%
    \fi
  \fi
    \setlength{\unitlength}{0.0500bp}%
    \ifx\gptboxheight\undefined%
      \newlength{\gptboxheight}%
      \newlength{\gptboxwidth}%
      \newsavebox{\gptboxtext}%
    \fi%
    \setlength{\fboxrule}{0.5pt}%
    \setlength{\fboxsep}{1pt}%
\begin{picture}(5668.00,4250.00)%
    \gplgaddtomacro\gplbacktext{%
      \csname LTb\endcsname%
      \put(946,704){\makebox(0,0)[r]{\strut{}$-1$}}%
      \put(946,1251){\makebox(0,0)[r]{\strut{}$-0.5$}}%
      \put(946,1798){\makebox(0,0)[r]{\strut{}$0$}}%
      \put(946,2345){\makebox(0,0)[r]{\strut{}$0.5$}}%
      \put(946,2891){\makebox(0,0)[r]{\strut{}$1$}}%
      \put(946,3438){\makebox(0,0)[r]{\strut{}$1.5$}}%
      \put(946,3985){\makebox(0,0)[r]{\strut{}$2$}}%
      \put(1078,484){\makebox(0,0){\strut{}$1$}}%
      \put(1917,484){\makebox(0,0){\strut{}$1.2$}}%
      \put(2755,484){\makebox(0,0){\strut{}$1.4$}}%
      \put(3594,484){\makebox(0,0){\strut{}$1.6$}}%
      \put(4432,484){\makebox(0,0){\strut{}$1.8$}}%
      \put(5271,484){\makebox(0,0){\strut{}$2$}}%
    }%
    \gplgaddtomacro\gplfronttext{%
      \csname LTb\endcsname%
      \put(176,2344){\rotatebox{-270}{\makebox(0,0){\strut{}${\cal G}_{\rm RN}$}}}%
      \put(3174,154){\makebox(0,0){\strut{}$r$}}%
      \csname LTb\endcsname%
      \put(4284,3812){\makebox(0,0)[r]{\strut{}$Q=\sqrt{2}$}}%
      \csname LTb\endcsname%
      \put(4284,3592){\makebox(0,0)[r]{\strut{}$Q=1.4$}}%
      \csname LTb\endcsname%
      \put(4284,3372){\makebox(0,0)[r]{\strut{}$Q=1.2$}}%
      \csname LTb\endcsname%
      \put(4284,3152){\makebox(0,0)[r]{\strut{}$Q=1.1$}}%
      \csname LTb\endcsname%
      \put(4284,2932){\makebox(0,0)[r]{\strut{}$Q=1.05$}}%
      \csname LTb\endcsname%
      \put(4284,2712){\makebox(0,0)[r]{\strut{}$Q=1$}}%
    }%
    \gplbacktext
    \put(0,0){\includegraphics{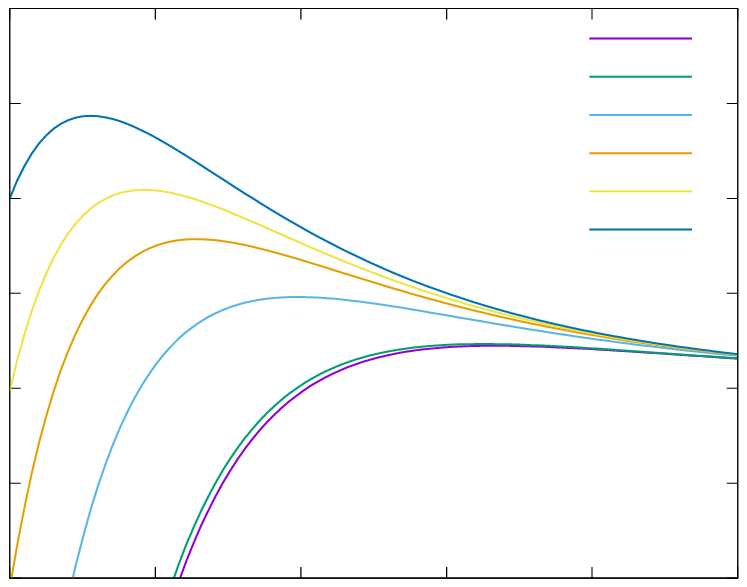}}%
    \gplfronttext
  \end{picture}%
\endgroup

%% file: profile_q14.tex
% GNUPLOT: LaTeX picture with Postscript
\begingroup
  \makeatletter
  \providecommand\color[2][]{%
    \GenericError{(gnuplot) \space\space\space\@spaces}{%
      Package color not loaded in conjunction with
      terminal option `colourtext'%
    }{See the gnuplot documentation for explanation.%
    }{Either use 'blacktext' in gnuplot or load the package
      color.sty in LaTeX.}%
    \renewcommand\color[2][]{}%
  }%
  \providecommand\includegraphics[2][]{%
    \GenericError{(gnuplot) \space\space\space\@spaces}{%
      Package graphicx or graphics not loaded%
    }{See the gnuplot documentation for explanation.%
    }{The gnuplot epslatex terminal needs graphicx.sty or graphics.sty.}%
    \renewcommand\includegraphics[2][]{}%
  }%
  \providecommand\rotatebox[2]{#2}%
  \@ifundefined{ifGPcolor}{%
    \newif\ifGPcolor
    \GPcolortrue
  }{}%
  \@ifundefined{ifGPblacktext}{%
    \newif\ifGPblacktext
    \GPblacktextfalse
  }{}%
  % define a \g@addto@macro without @ in the name:
  \let\gplgaddtomacro\g@addto@macro
  % define empty templates for all commands taking text:
  \gdef\gplbacktext{}%
  \gdef\gplfronttext{}%
  \makeatother
  \ifGPblacktext
    % no textcolor at all
    \def\colorrgb#1{}%
    \def\colorgray#1{}%
  \else
    % gray or color?
    \ifGPcolor
      \def\colorrgb#1{\color[rgb]{#1}}%
      \def\colorgray#1{\color[gray]{#1}}%
      \expandafter\def\csname LTw\endcsname{\color{white}}%
      \expandafter\def\csname LTb\endcsname{\color{black}}%
      \expandafter\def\csname LTa\endcsname{\color{black}}%
      \expandafter\def\csname LT0\endcsname{\color[rgb]{1,0,0}}%
      \expandafter\def\csname LT1\endcsname{\color[rgb]{0,1,0}}%
      \expandafter\def\csname LT2\endcsname{\color[rgb]{0,0,1}}%
      \expandafter\def\csname LT3\endcsname{\color[rgb]{1,0,1}}%
      \expandafter\def\csname LT4\endcsname{\color[rgb]{0,1,1}}%
      \expandafter\def\csname LT5\endcsname{\color[rgb]{1,1,0}}%
      \expandafter\def\csname LT6\endcsname{\color[rgb]{0,0,0}}%
      \expandafter\def\csname LT7\endcsname{\color[rgb]{1,0.3,0}}%
      \expandafter\def\csname LT8\endcsname{\color[rgb]{0.5,0.5,0.5}}%
    \else
      % gray
      \def\colorrgb#1{\color{black}}%
      \def\colorgray#1{\color[gray]{#1}}%
      \expandafter\def\csname LTw\endcsname{\color{white}}%
      \expandafter\def\csname LTb\endcsname{\color{black}}%
      \expandafter\def\csname LTa\endcsname{\color{black}}%
      \expandafter\def\csname LT0\endcsname{\color{black}}%
      \expandafter\def\csname LT1\endcsname{\color{black}}%
      \expandafter\def\csname LT2\endcsname{\color{black}}%
      \expandafter\def\csname LT3\endcsname{\color{black}}%
      \expandafter\def\csname LT4\endcsname{\color{black}}%
      \expandafter\def\csname LT5\endcsname{\color{black}}%
      \expandafter\def\csname LT6\endcsname{\color{black}}%
      \expandafter\def\csname LT7\endcsname{\color{black}}%
      \expandafter\def\csname LT8\endcsname{\color{black}}%
    \fi
  \fi
    \setlength{\unitlength}{0.0500bp}%
    \ifx\gptboxheight\undefined%
      \newlength{\gptboxheight}%
      \newlength{\gptboxwidth}%
      \newsavebox{\gptboxtext}%
    \fi%
    \setlength{\fboxrule}{0.5pt}%
    \setlength{\fboxsep}{1pt}%
\begin{picture}(5668.00,4250.00)%
    \gplgaddtomacro\gplbacktext{%
      \csname LTb\endcsname%
      \put(1210,704){\makebox(0,0)[r]{\strut{}$0.001$}}%
      \put(1210,1114){\makebox(0,0)[r]{\strut{}$0.01$}}%
      \put(1210,1524){\makebox(0,0)[r]{\strut{}$0.1$}}%
      \put(1210,1934){\makebox(0,0)[r]{\strut{}$1$}}%
      \put(1210,2345){\makebox(0,0)[r]{\strut{}$10$}}%
      \put(1210,2755){\makebox(0,0)[r]{\strut{}$100$}}%
      \put(1210,3165){\makebox(0,0)[r]{\strut{}$1000$}}%
      \put(1210,3575){\makebox(0,0)[r]{\strut{}$10000$}}%
      \put(1210,3985){\makebox(0,0)[r]{\strut{}$100000$}}%
      \put(1342,484){\makebox(0,0){\strut{}$1$}}%
      \put(2128,484){\makebox(0,0){\strut{}$10$}}%
      \put(2914,484){\makebox(0,0){\strut{}$100$}}%
      \put(3699,484){\makebox(0,0){\strut{}$1000$}}%
      \put(4485,484){\makebox(0,0){\strut{}$10000$}}%
      \put(5271,484){\makebox(0,0){\strut{}$100000$}}%
    }%
    \gplgaddtomacro\gplfronttext{%
      \csname LTb\endcsname%
      \put(176,2344){\rotatebox{-270}{\makebox(0,0){\strut{}$\phi(r)$}}}%
      \put(3306,154){\makebox(0,0){\strut{}$r$}}%
      \csname LTb\endcsname%
      \put(4284,3812){\makebox(0,0)[r]{\strut{}$\gamma=1.954$}}%
      \csname LTb\endcsname%
      \put(4284,3592){\makebox(0,0)[r]{\strut{}$\gamma=-0.082$}}%
    }%
    \gplbacktext
    \put(0,0){\includegraphics{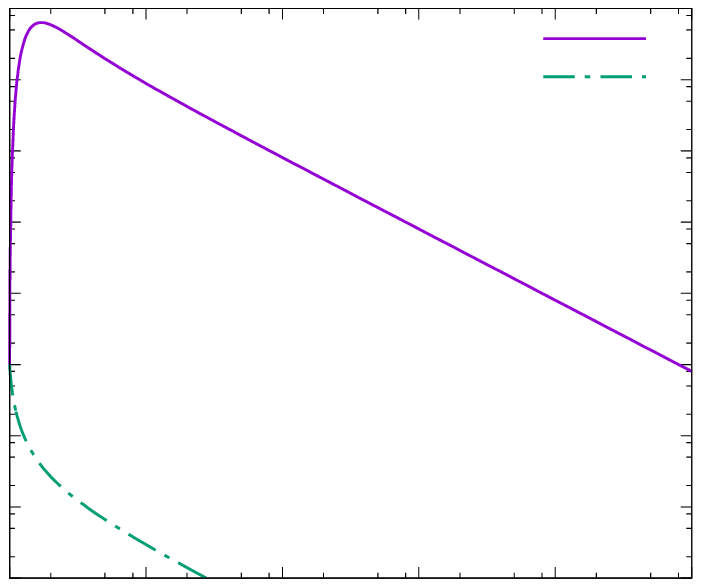}}%
    \gplfronttext
  \end{picture}%
\endgroup

%% file: domain_q_ga.tex
% GNUPLOT: LaTeX picture with Postscript
\begingroup
  \makeatletter
  \providecommand\color[2][]{%
    \GenericError{(gnuplot) \space\space\space\@spaces}{%
      Package color not loaded in conjunction with
      terminal option `colourtext'%
    }{See the gnuplot documentation for explanation.%
    }{Either use 'blacktext' in gnuplot or load the package
      color.sty in LaTeX.}%
    \renewcommand\color[2][]{}%
  }%
  \providecommand\includegraphics[2][]{%
    \GenericError{(gnuplot) \space\space\space\@spaces}{%
      Package graphicx or graphics not loaded%
    }{See the gnuplot documentation for explanation.%
    }{The gnuplot epslatex terminal needs graphicx.sty or graphics.sty.}%
    \renewcommand\includegraphics[2][]{}%
  }%
  \providecommand\rotatebox[2]{#2}%
  \@ifundefined{ifGPcolor}{%
    \newif\ifGPcolor
    \GPcolortrue
  }{}%
  \@ifundefined{ifGPblacktext}{%
    \newif\ifGPblacktext
    \GPblacktextfalse
  }{}%
  % define a \g@addto@macro without @ in the name:
  \let\gplgaddtomacro\g@addto@macro
  % define empty templates for all commands taking text:
  \gdef\gplbacktext{}%
  \gdef\gplfronttext{}%
  \makeatother
  \ifGPblacktext
    % no textcolor at all
    \def\colorrgb#1{}%
    \def\colorgray#1{}%
  \else
    % gray or color?
    \ifGPcolor
      \def\colorrgb#1{\color[rgb]{#1}}%
      \def\colorgray#1{\color[gray]{#1}}%
      \expandafter\def\csname LTw\endcsname{\color{white}}%
      \expandafter\def\csname LTb\endcsname{\color{black}}%
      \expandafter\def\csname LTa\endcsname{\color{black}}%
      \expandafter\def\csname LT0\endcsname{\color[rgb]{1,0,0}}%
      \expandafter\def\csname LT1\endcsname{\color[rgb]{0,1,0}}%
      \expandafter\def\csname LT2\endcsname{\color[rgb]{0,0,1}}%
      \expandafter\def\csname LT3\endcsname{\color[rgb]{1,0,1}}%
      \expandafter\def\csname LT4\endcsname{\color[rgb]{0,1,1}}%
      \expandafter\def\csname LT5\endcsname{\color[rgb]{1,1,0}}%
      \expandafter\def\csname LT6\endcsname{\color[rgb]{0,0,0}}%
      \expandafter\def\csname LT7\endcsname{\color[rgb]{1,0.3,0}}%
      \expandafter\def\csname LT8\endcsname{\color[rgb]{0.5,0.5,0.5}}%
    \else
      % gray
      \def\colorrgb#1{\color{black}}%
      \def\colorgray#1{\color[gray]{#1}}%
      \expandafter\def\csname LTw\endcsname{\color{white}}%
      \expandafter\def\csname LTb\endcsname{\color{black}}%
      \expandafter\def\csname LTa\endcsname{\color{black}}%
      \expandafter\def\csname LT0\endcsname{\color{black}}%
      \expandafter\def\csname LT1\endcsname{\color{black}}%
      \expandafter\def\csname LT2\endcsname{\color{black}}%
      \expandafter\def\csname LT3\endcsname{\color{black}}%
      \expandafter\def\csname LT4\endcsname{\color{black}}%
      \expandafter\def\csname LT5\endcsname{\color{black}}%
      \expandafter\def\csname LT6\endcsname{\color{black}}%
      \expandafter\def\csname LT7\endcsname{\color{black}}%
      \expandafter\def\csname LT8\endcsname{\color{black}}%
    \fi
  \fi
    \setlength{\unitlength}{0.0500bp}%
    \ifx\gptboxheight\undefined%
      \newlength{\gptboxheight}%
      \newlength{\gptboxwidth}%
      \newsavebox{\gptboxtext}%
    \fi%
    \setlength{\fboxrule}{0.5pt}%
    \setlength{\fboxsep}{1pt}%
\begin{picture}(5668.00,4250.00)%
    \gplgaddtomacro\gplbacktext{%
      \csname LTb\endcsname%
      \put(946,704){\makebox(0,0)[r]{\strut{}$-1.5$}}%
      \put(946,1173){\makebox(0,0)[r]{\strut{}$-1$}}%
      \put(946,1641){\makebox(0,0)[r]{\strut{}$-0.5$}}%
      \put(946,2110){\makebox(0,0)[r]{\strut{}$0$}}%
      \put(946,2579){\makebox(0,0)[r]{\strut{}$0.5$}}%
      \put(946,3048){\makebox(0,0)[r]{\strut{}$1$}}%
      \put(946,3516){\makebox(0,0)[r]{\strut{}$1.5$}}%
      \put(946,3985){\makebox(0,0)[r]{\strut{}$2$}}%
      \put(1078,484){\makebox(0,0){\strut{}$0$}}%
      \put(1917,484){\makebox(0,0){\strut{}$0.3$}}%
      \put(2755,484){\makebox(0,0){\strut{}$0.6$}}%
      \put(3594,484){\makebox(0,0){\strut{}$0.9$}}%
      \put(4432,484){\makebox(0,0){\strut{}$1.2$}}%
      \put(5271,484){\makebox(0,0){\strut{}$1.5$}}%
      \put(1777,3516){\makebox(0,0)[l]{\strut{}$\gamma_{0}$}}%
      \put(1777,3048){\makebox(0,0)[l]{\strut{}$\gamma_{\rm cr}$}}%
    }%
    \gplgaddtomacro\gplfronttext{%
      \csname LTb\endcsname%
      \put(176,2344){\rotatebox{-270}{\makebox(0,0){\strut{}$\gamma$}}}%
      \put(3174,154){\makebox(0,0){\strut{}$Q$}}%
    }%
    \gplbacktext
    \put(0,0){\includegraphics{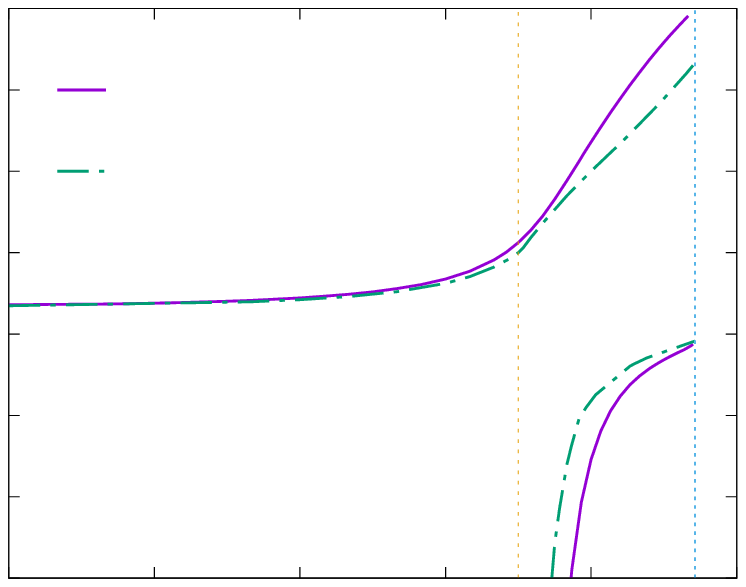}}%
    \gplfronttext
  \end{picture}%
\endgroup

%% file: profile_N_phi0_001.tex
% GNUPLOT: LaTeX picture with Postscript
\begingroup
  \makeatletter
  \providecommand\color[2][]{%
    \GenericError{(gnuplot) \space\space\space\@spaces}{%
      Package color not loaded in conjunction with
      terminal option `colourtext'%
    }{See the gnuplot documentation for explanation.%
    }{Either use 'blacktext' in gnuplot or load the package
      color.sty in LaTeX.}%
    \renewcommand\color[2][]{}%
  }%
  \providecommand\includegraphics[2][]{%
    \GenericError{(gnuplot) \space\space\space\@spaces}{%
      Package graphicx or graphics not loaded%
    }{See the gnuplot documentation for explanation.%
    }{The gnuplot epslatex terminal needs graphicx.sty or graphics.sty.}%
    \renewcommand\includegraphics[2][]{}%
  }%
  \providecommand\rotatebox[2]{#2}%
  \@ifundefined{ifGPcolor}{%
    \newif\ifGPcolor
    \GPcolortrue
  }{}%
  \@ifundefined{ifGPblacktext}{%
    \newif\ifGPblacktext
    \GPblacktextfalse
  }{}%
  % define a \g@addto@macro without @ in the name:
  \let\gplgaddtomacro\g@addto@macro
  % define empty templates for all commands taking text:
  \gdef\gplbacktext{}%
  \gdef\gplfronttext{}%
  \makeatother
  \ifGPblacktext
    % no textcolor at all
    \def\colorrgb#1{}%
    \def\colorgray#1{}%
  \else
    % gray or color?
    \ifGPcolor
      \def\colorrgb#1{\color[rgb]{#1}}%
      \def\colorgray#1{\color[gray]{#1}}%
      \expandafter\def\csname LTw\endcsname{\color{white}}%
      \expandafter\def\csname LTb\endcsname{\color{black}}%
      \expandafter\def\csname LTa\endcsname{\color{black}}%
      \expandafter\def\csname LT0\endcsname{\color[rgb]{1,0,0}}%
      \expandafter\def\csname LT1\endcsname{\color[rgb]{0,1,0}}%
      \expandafter\def\csname LT2\endcsname{\color[rgb]{0,0,1}}%
      \expandafter\def\csname LT3\endcsname{\color[rgb]{1,0,1}}%
      \expandafter\def\csname LT4\endcsname{\color[rgb]{0,1,1}}%
      \expandafter\def\csname LT5\endcsname{\color[rgb]{1,1,0}}%
      \expandafter\def\csname LT6\endcsname{\color[rgb]{0,0,0}}%
      \expandafter\def\csname LT7\endcsname{\color[rgb]{1,0.3,0}}%
      \expandafter\def\csname LT8\endcsname{\color[rgb]{0.5,0.5,0.5}}%
    \else
      % gray
      \def\colorrgb#1{\color{black}}%
      \def\colorgray#1{\color[gray]{#1}}%
      \expandafter\def\csname LTw\endcsname{\color{white}}%
      \expandafter\def\csname LTb\endcsname{\color{black}}%
      \expandafter\def\csname LTa\endcsname{\color{black}}%
      \expandafter\def\csname LT0\endcsname{\color{black}}%
      \expandafter\def\csname LT1\endcsname{\color{black}}%
      \expandafter\def\csname LT2\endcsname{\color{black}}%
      \expandafter\def\csname LT3\endcsname{\color{black}}%
      \expandafter\def\csname LT4\endcsname{\color{black}}%
      \expandafter\def\csname LT5\endcsname{\color{black}}%
      \expandafter\def\csname LT6\endcsname{\color{black}}%
      \expandafter\def\csname LT7\endcsname{\color{black}}%
      \expandafter\def\csname LT8\endcsname{\color{black}}%
    \fi
  \fi
    \setlength{\unitlength}{0.0500bp}%
    \ifx\gptboxheight\undefined%
      \newlength{\gptboxheight}%
      \newlength{\gptboxwidth}%
      \newsavebox{\gptboxtext}%
    \fi%
    \setlength{\fboxrule}{0.5pt}%
    \setlength{\fboxsep}{1pt}%
\begin{picture}(5102.00,4250.00)%
    \gplgaddtomacro\gplbacktext{%
      \csname LTb\endcsname%
      \put(946,704){\makebox(0,0)[r]{\strut{}$0$}}%
      \put(946,1360){\makebox(0,0)[r]{\strut{}$0.04$}}%
      \put(946,2016){\makebox(0,0)[r]{\strut{}$0.08$}}%
      \put(946,2673){\makebox(0,0)[r]{\strut{}$0.12$}}%
      \put(946,3329){\makebox(0,0)[r]{\strut{}$0.16$}}%
      \put(946,3985){\makebox(0,0)[r]{\strut{}$0.2$}}%
      \put(1078,484){\makebox(0,0){\strut{}$1$}}%
      \put(1803,484){\makebox(0,0){\strut{}$1.1$}}%
      \put(2529,484){\makebox(0,0){\strut{}$1.2$}}%
      \put(3254,484){\makebox(0,0){\strut{}$1.3$}}%
      \put(3980,484){\makebox(0,0){\strut{}$1.4$}}%
      \put(4705,484){\makebox(0,0){\strut{}$1.5$}}%
    }%
    \gplgaddtomacro\gplfronttext{%
      \csname LTb\endcsname%
      \put(176,2344){\rotatebox{-270}{\makebox(0,0){\strut{}$N(r)$}}}%
      \put(2891,154){\makebox(0,0){\strut{}$r$}}%
      \csname LTb\endcsname%
      \put(2398,3812){\makebox(0,0)[r]{\strut{}$Q=1.2$}}%
      \csname LTb\endcsname%
      \put(2398,3592){\makebox(0,0)[r]{\strut{}$Q=1.24$}}%
      \csname LTb\endcsname%
      \put(2398,3372){\makebox(0,0)[r]{\strut{}$Q=1.26$}}%
      \csname LTb\endcsname%
      \put(2398,3152){\makebox(0,0)[r]{\strut{}$Q=1.264$}}%
      \csname LTb\endcsname%
      \put(2398,2932){\makebox(0,0)[r]{\strut{}$Q=1.26685$}}%
    }%
    \gplbacktext
    \put(0,0){\includegraphics{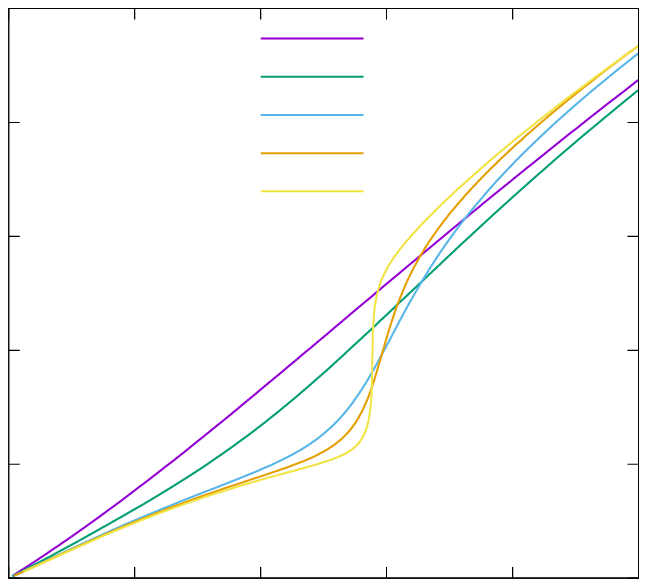}}%
    \gplfronttext
  \end{picture}%
\endgroup

%% file: profile_kretsch_phi0_001.tex
% GNUPLOT: LaTeX picture with Postscript
\begingroup
  \makeatletter
  \providecommand\color[2][]{%
    \GenericError{(gnuplot) \space\space\space\@spaces}{%
      Package color not loaded in conjunction with
      terminal option `colourtext'%
    }{See the gnuplot documentation for explanation.%
    }{Either use 'blacktext' in gnuplot or load the package
      color.sty in LaTeX.}%
    \renewcommand\color[2][]{}%
  }%
  \providecommand\includegraphics[2][]{%
    \GenericError{(gnuplot) \space\space\space\@spaces}{%
      Package graphicx or graphics not loaded%
    }{See the gnuplot documentation for explanation.%
    }{The gnuplot epslatex terminal needs graphicx.sty or graphics.sty.}%
    \renewcommand\includegraphics[2][]{}%
  }%
  \providecommand\rotatebox[2]{#2}%
  \@ifundefined{ifGPcolor}{%
    \newif\ifGPcolor
    \GPcolortrue
  }{}%
  \@ifundefined{ifGPblacktext}{%
    \newif\ifGPblacktext
    \GPblacktextfalse
  }{}%
  % define a \g@addto@macro without @ in the name:
  \let\gplgaddtomacro\g@addto@macro
  % define empty templates for all commands taking text:
  \gdef\gplbacktext{}%
  \gdef\gplfronttext{}%
  \makeatother
  \ifGPblacktext
    % no textcolor at all
    \def\colorrgb#1{}%
    \def\colorgray#1{}%
  \else
    % gray or color?
    \ifGPcolor
      \def\colorrgb#1{\color[rgb]{#1}}%
      \def\colorgray#1{\color[gray]{#1}}%
      \expandafter\def\csname LTw\endcsname{\color{white}}%
      \expandafter\def\csname LTb\endcsname{\color{black}}%
      \expandafter\def\csname LTa\endcsname{\color{black}}%
      \expandafter\def\csname LT0\endcsname{\color[rgb]{1,0,0}}%
      \expandafter\def\csname LT1\endcsname{\color[rgb]{0,1,0}}%
      \expandafter\def\csname LT2\endcsname{\color[rgb]{0,0,1}}%
      \expandafter\def\csname LT3\endcsname{\color[rgb]{1,0,1}}%
      \expandafter\def\csname LT4\endcsname{\color[rgb]{0,1,1}}%
      \expandafter\def\csname LT5\endcsname{\color[rgb]{1,1,0}}%
      \expandafter\def\csname LT6\endcsname{\color[rgb]{0,0,0}}%
      \expandafter\def\csname LT7\endcsname{\color[rgb]{1,0.3,0}}%
      \expandafter\def\csname LT8\endcsname{\color[rgb]{0.5,0.5,0.5}}%
    \else
      % gray
      \def\colorrgb#1{\color{black}}%
      \def\colorgray#1{\color[gray]{#1}}%
      \expandafter\def\csname LTw\endcsname{\color{white}}%
      \expandafter\def\csname LTb\endcsname{\color{black}}%
      \expandafter\def\csname LTa\endcsname{\color{black}}%
      \expandafter\def\csname LT0\endcsname{\color{black}}%
      \expandafter\def\csname LT1\endcsname{\color{black}}%
      \expandafter\def\csname LT2\endcsname{\color{black}}%
      \expandafter\def\csname LT3\endcsname{\color{black}}%
      \expandafter\def\csname LT4\endcsname{\color{black}}%
      \expandafter\def\csname LT5\endcsname{\color{black}}%
      \expandafter\def\csname LT6\endcsname{\color{black}}%
      \expandafter\def\csname LT7\endcsname{\color{black}}%
      \expandafter\def\csname LT8\endcsname{\color{black}}%
    \fi
  \fi
    \setlength{\unitlength}{0.0500bp}%
    \ifx\gptboxheight\undefined%
      \newlength{\gptboxheight}%
      \newlength{\gptboxwidth}%
      \newsavebox{\gptboxtext}%
    \fi%
    \setlength{\fboxrule}{0.5pt}%
    \setlength{\fboxsep}{1pt}%
\begin{picture}(5102.00,4250.00)%
    \gplgaddtomacro\gplbacktext{%
      \csname LTb\endcsname%
      \put(946,704){\makebox(0,0)[r]{\strut{}$0.01$}}%
      \put(946,1798){\makebox(0,0)[r]{\strut{}$0.1$}}%
      \put(946,2891){\makebox(0,0)[r]{\strut{}$1$}}%
      \put(946,3985){\makebox(0,0)[r]{\strut{}$10$}}%
      \put(1078,484){\makebox(0,0){\strut{}$1$}}%
      \put(1803,484){\makebox(0,0){\strut{}$1.1$}}%
      \put(2529,484){\makebox(0,0){\strut{}$1.2$}}%
      \put(3254,484){\makebox(0,0){\strut{}$1.3$}}%
      \put(3980,484){\makebox(0,0){\strut{}$1.4$}}%
      \put(4705,484){\makebox(0,0){\strut{}$1.5$}}%
    }%
    \gplgaddtomacro\gplfronttext{%
      \csname LTb\endcsname%
      \put(176,2344){\rotatebox{-270}{\makebox(0,0){\strut{}$K/24$}}}%
      \put(2891,154){\makebox(0,0){\strut{}$r$}}%
    }%
    \gplbacktext
    \put(0,0){\includegraphics{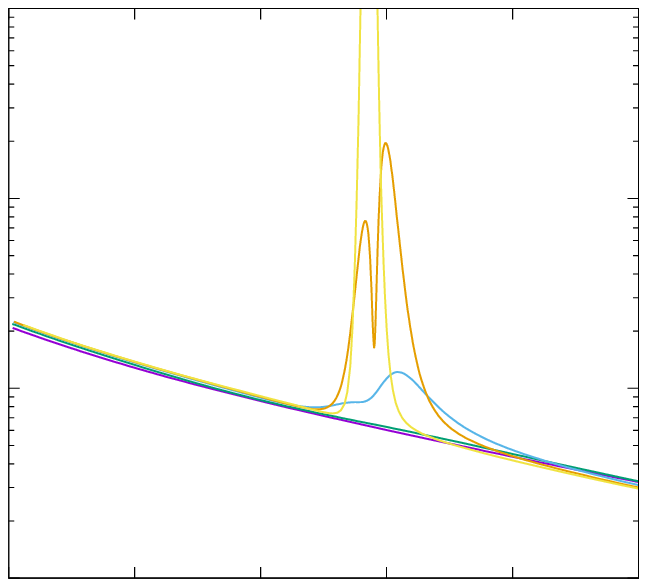}}%
    \gplfronttext
  \end{picture}%
\endgroup

%% file: profile_vp_phi0_001.tex
% GNUPLOT: LaTeX picture with Postscript
\begingroup
  \makeatletter
  \providecommand\color[2][]{%
    \GenericError{(gnuplot) \space\space\space\@spaces}{%
      Package color not loaded in conjunction with
      terminal option `colourtext'%
    }{See the gnuplot documentation for explanation.%
    }{Either use 'blacktext' in gnuplot or load the package
      color.sty in LaTeX.}%
    \renewcommand\color[2][]{}%
  }%
  \providecommand\includegraphics[2][]{%
    \GenericError{(gnuplot) \space\space\space\@spaces}{%
      Package graphicx or graphics not loaded%
    }{See the gnuplot documentation for explanation.%
    }{The gnuplot epslatex terminal needs graphicx.sty or graphics.sty.}%
    \renewcommand\includegraphics[2][]{}%
  }%
  \providecommand\rotatebox[2]{#2}%
  \@ifundefined{ifGPcolor}{%
    \newif\ifGPcolor
    \GPcolortrue
  }{}%
  \@ifundefined{ifGPblacktext}{%
    \newif\ifGPblacktext
    \GPblacktextfalse
  }{}%
  % define a \g@addto@macro without @ in the name:
  \let\gplgaddtomacro\g@addto@macro
  % define empty templates for all commands taking text:
  \gdef\gplbacktext{}%
  \gdef\gplfronttext{}%
  \makeatother
  \ifGPblacktext
    % no textcolor at all
    \def\colorrgb#1{}%
    \def\colorgray#1{}%
  \else
    % gray or color?
    \ifGPcolor
      \def\colorrgb#1{\color[rgb]{#1}}%
      \def\colorgray#1{\color[gray]{#1}}%
      \expandafter\def\csname LTw\endcsname{\color{white}}%
      \expandafter\def\csname LTb\endcsname{\color{black}}%
      \expandafter\def\csname LTa\endcsname{\color{black}}%
      \expandafter\def\csname LT0\endcsname{\color[rgb]{1,0,0}}%
      \expandafter\def\csname LT1\endcsname{\color[rgb]{0,1,0}}%
      \expandafter\def\csname LT2\endcsname{\color[rgb]{0,0,1}}%
      \expandafter\def\csname LT3\endcsname{\color[rgb]{1,0,1}}%
      \expandafter\def\csname LT4\endcsname{\color[rgb]{0,1,1}}%
      \expandafter\def\csname LT5\endcsname{\color[rgb]{1,1,0}}%
      \expandafter\def\csname LT6\endcsname{\color[rgb]{0,0,0}}%
      \expandafter\def\csname LT7\endcsname{\color[rgb]{1,0.3,0}}%
      \expandafter\def\csname LT8\endcsname{\color[rgb]{0.5,0.5,0.5}}%
    \else
      % gray
      \def\colorrgb#1{\color{black}}%
      \def\colorgray#1{\color[gray]{#1}}%
      \expandafter\def\csname LTw\endcsname{\color{white}}%
      \expandafter\def\csname LTb\endcsname{\color{black}}%
      \expandafter\def\csname LTa\endcsname{\color{black}}%
      \expandafter\def\csname LT0\endcsname{\color{black}}%
      \expandafter\def\csname LT1\endcsname{\color{black}}%
      \expandafter\def\csname LT2\endcsname{\color{black}}%
      \expandafter\def\csname LT3\endcsname{\color{black}}%
      \expandafter\def\csname LT4\endcsname{\color{black}}%
      \expandafter\def\csname LT5\endcsname{\color{black}}%
      \expandafter\def\csname LT6\endcsname{\color{black}}%
      \expandafter\def\csname LT7\endcsname{\color{black}}%
      \expandafter\def\csname LT8\endcsname{\color{black}}%
    \fi
  \fi
    \setlength{\unitlength}{0.0500bp}%
    \ifx\gptboxheight\undefined%
      \newlength{\gptboxheight}%
      \newlength{\gptboxwidth}%
      \newsavebox{\gptboxtext}%
    \fi%
    \setlength{\fboxrule}{0.5pt}%
    \setlength{\fboxsep}{1pt}%
\begin{picture}(5102.00,4250.00)%
    \gplgaddtomacro\gplbacktext{%
      \csname LTb\endcsname%
      \put(946,704){\makebox(0,0)[r]{\strut{}$0.25$}}%
      \put(946,1524){\makebox(0,0)[r]{\strut{}$0.5$}}%
      \put(946,2345){\makebox(0,0)[r]{\strut{}$0.75$}}%
      \put(946,3165){\makebox(0,0)[r]{\strut{}$1$}}%
      \put(946,3985){\makebox(0,0)[r]{\strut{}$1.25$}}%
      \put(1078,484){\makebox(0,0){\strut{}$1$}}%
      \put(1803,484){\makebox(0,0){\strut{}$1.1$}}%
      \put(2529,484){\makebox(0,0){\strut{}$1.2$}}%
      \put(3254,484){\makebox(0,0){\strut{}$1.3$}}%
      \put(3980,484){\makebox(0,0){\strut{}$1.4$}}%
      \put(4705,484){\makebox(0,0){\strut{}$1.5$}}%
    }%
    \gplgaddtomacro\gplfronttext{%
      \csname LTb\endcsname%
      \put(176,2344){\rotatebox{-270}{\makebox(0,0){\strut{}$\frac{dV}{dr}$}}}%
      \put(2891,154){\makebox(0,0){\strut{}$r$}}%
      \csname LTb\endcsname%
      \put(2398,1757){\makebox(0,0)[r]{\strut{}$Q=1.2$}}%
      \csname LTb\endcsname%
      \put(2398,1537){\makebox(0,0)[r]{\strut{}$Q=1.24$}}%
      \csname LTb\endcsname%
      \put(2398,1317){\makebox(0,0)[r]{\strut{}$Q=1.26$}}%
      \csname LTb\endcsname%
      \put(2398,1097){\makebox(0,0)[r]{\strut{}$Q=1.264$}}%
      \csname LTb\endcsname%
      \put(2398,877){\makebox(0,0)[r]{\strut{}$Q=1.26685$}}%
    }%
    \gplbacktext
    \put(0,0){\includegraphics{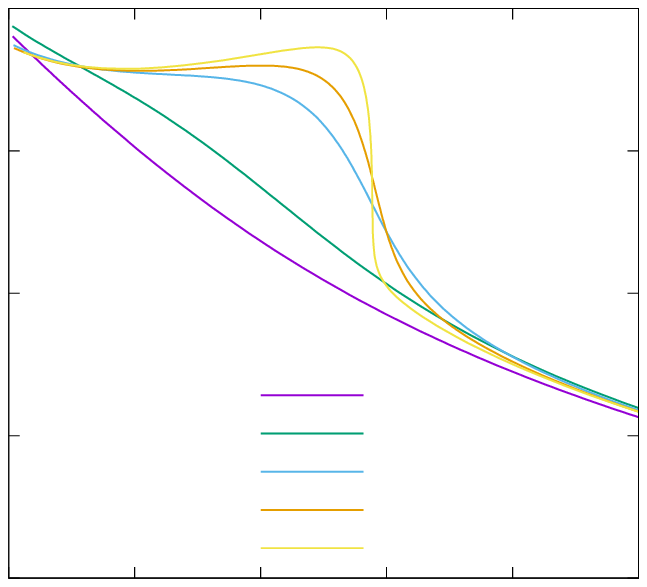}}%
    \gplfronttext
  \end{picture}%
\endgroup

%% file: profile_phi_phi0_001.tex
% GNUPLOT: LaTeX picture with Postscript
\begingroup
  \makeatletter
  \providecommand\color[2][]{%
    \GenericError{(gnuplot) \space\space\space\@spaces}{%
      Package color not loaded in conjunction with
      terminal option `colourtext'%
    }{See the gnuplot documentation for explanation.%
    }{Either use 'blacktext' in gnuplot or load the package
      color.sty in LaTeX.}%
    \renewcommand\color[2][]{}%
  }%
  \providecommand\includegraphics[2][]{%
    \GenericError{(gnuplot) \space\space\space\@spaces}{%
      Package graphicx or graphics not loaded%
    }{See the gnuplot documentation for explanation.%
    }{The gnuplot epslatex terminal needs graphicx.sty or graphics.sty.}%
    \renewcommand\includegraphics[2][]{}%
  }%
  \providecommand\rotatebox[2]{#2}%
  \@ifundefined{ifGPcolor}{%
    \newif\ifGPcolor
    \GPcolortrue
  }{}%
  \@ifundefined{ifGPblacktext}{%
    \newif\ifGPblacktext
    \GPblacktextfalse
  }{}%
  % define a \g@addto@macro without @ in the name:
  \let\gplgaddtomacro\g@addto@macro
  % define empty templates for all commands taking text:
  \gdef\gplbacktext{}%
  \gdef\gplfronttext{}%
  \makeatother
  \ifGPblacktext
    % no textcolor at all
    \def\colorrgb#1{}%
    \def\colorgray#1{}%
  \else
    % gray or color?
    \ifGPcolor
      \def\colorrgb#1{\color[rgb]{#1}}%
      \def\colorgray#1{\color[gray]{#1}}%
      \expandafter\def\csname LTw\endcsname{\color{white}}%
      \expandafter\def\csname LTb\endcsname{\color{black}}%
      \expandafter\def\csname LTa\endcsname{\color{black}}%
      \expandafter\def\csname LT0\endcsname{\color[rgb]{1,0,0}}%
      \expandafter\def\csname LT1\endcsname{\color[rgb]{0,1,0}}%
      \expandafter\def\csname LT2\endcsname{\color[rgb]{0,0,1}}%
      \expandafter\def\csname LT3\endcsname{\color[rgb]{1,0,1}}%
      \expandafter\def\csname LT4\endcsname{\color[rgb]{0,1,1}}%
      \expandafter\def\csname LT5\endcsname{\color[rgb]{1,1,0}}%
      \expandafter\def\csname LT6\endcsname{\color[rgb]{0,0,0}}%
      \expandafter\def\csname LT7\endcsname{\color[rgb]{1,0.3,0}}%
      \expandafter\def\csname LT8\endcsname{\color[rgb]{0.5,0.5,0.5}}%
    \else
      % gray
      \def\colorrgb#1{\color{black}}%
      \def\colorgray#1{\color[gray]{#1}}%
      \expandafter\def\csname LTw\endcsname{\color{white}}%
      \expandafter\def\csname LTb\endcsname{\color{black}}%
      \expandafter\def\csname LTa\endcsname{\color{black}}%
      \expandafter\def\csname LT0\endcsname{\color{black}}%
      \expandafter\def\csname LT1\endcsname{\color{black}}%
      \expandafter\def\csname LT2\endcsname{\color{black}}%
      \expandafter\def\csname LT3\endcsname{\color{black}}%
      \expandafter\def\csname LT4\endcsname{\color{black}}%
      \expandafter\def\csname LT5\endcsname{\color{black}}%
      \expandafter\def\csname LT6\endcsname{\color{black}}%
      \expandafter\def\csname LT7\endcsname{\color{black}}%
      \expandafter\def\csname LT8\endcsname{\color{black}}%
    \fi
  \fi
    \setlength{\unitlength}{0.0500bp}%
    \ifx\gptboxheight\undefined%
      \newlength{\gptboxheight}%
      \newlength{\gptboxwidth}%
      \newsavebox{\gptboxtext}%
    \fi%
    \setlength{\fboxrule}{0.5pt}%
    \setlength{\fboxsep}{1pt}%
\begin{picture}(5102.00,4250.00)%
    \gplgaddtomacro\gplbacktext{%
      \csname LTb\endcsname%
      \put(946,704){\makebox(0,0)[r]{\strut{}$0$}}%
      \put(946,1524){\makebox(0,0)[r]{\strut{}$0.04$}}%
      \put(946,2345){\makebox(0,0)[r]{\strut{}$0.08$}}%
      \put(946,3165){\makebox(0,0)[r]{\strut{}$0.12$}}%
      \put(946,3985){\makebox(0,0)[r]{\strut{}$0.16$}}%
      \put(1078,484){\makebox(0,0){\strut{}$1$}}%
      \put(1803,484){\makebox(0,0){\strut{}$1.1$}}%
      \put(2529,484){\makebox(0,0){\strut{}$1.2$}}%
      \put(3254,484){\makebox(0,0){\strut{}$1.3$}}%
      \put(3980,484){\makebox(0,0){\strut{}$1.4$}}%
      \put(4705,484){\makebox(0,0){\strut{}$1.5$}}%
    }%
    \gplgaddtomacro\gplfronttext{%
      \csname LTb\endcsname%
      \put(176,2344){\rotatebox{-270}{\makebox(0,0){\strut{}$\phi(r)$}}}%
      \put(2891,154){\makebox(0,0){\strut{}$r$}}%
    }%
    \gplbacktext
    \put(0,0){\includegraphics{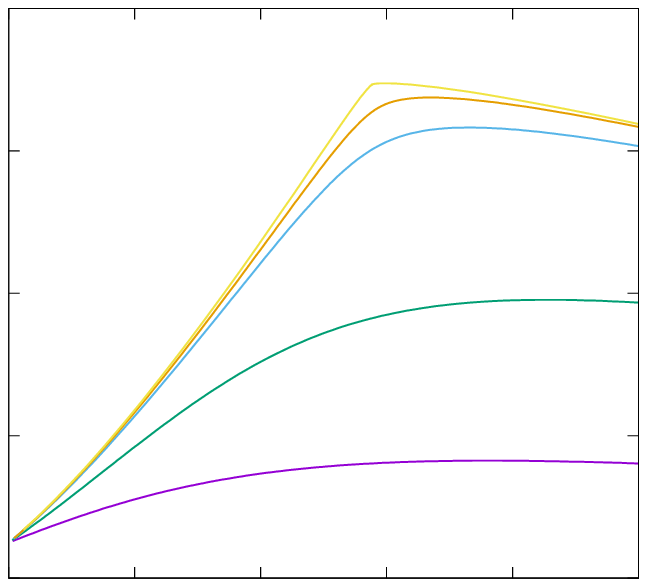}}%
    \gplfronttext
  \end{picture}%
\endgroup